\documentclass[apj]{emulateapj}
\usepackage{graphics}

\newcommand{\oiw}{\mbox{[\ion{O}{1}] $\lambda$6300}}
\newcommand{\oiiw}{\mbox{[\ion{O}{2}] $\lambda$3727}}

\newcommand{\niibw}{\mbox{[\ion{N}{2}] $\lambda$6583}}
\newcommand{\niiw}{\mbox{[\ion{N}{2}] $\lambda \lambda$6548,6583}}

\newcommand{\oiiibw}{\mbox{[\ion{O}{3}] $\lambda$5007}}
\newcommand{\oiiiw}{\mbox{[\ion{O}{3}] $\lambda \lambda$4959,5007}}

\newcommand{\oi}{\mbox{[\ion{O}{1}]}}
\newcommand{\oii}{\mbox{[\ion{O}{2}]}}
\newcommand{\nii}{\mbox{[\ion{N}{2}]}}
\newcommand{\hal}{\mbox{H$\alpha$}}

\newcommand{\hb}{\mbox{H$\beta$}}
\newcommand{\oiii}{\mbox{[\ion{O}{3}]}}

\shorttitle{[OII] in Red Sequence and Post-starburst Galaxies}
\shortauthors{Yan et al.}

\begin{document}
\title{On the Origin of [OII] Emission in Red Sequence and Post-starburst Galaxies}
\author{Renbin Yan$^{1}$, Jeffrey A. Newman$^{2,3}$, S.M. Faber$^{4}$, Nicholas Konidaris$^{4}$, David Koo$^4$, Marc Davis$^{1,5}$}
\affil{$^1$ Department of Astronomy, University of California, Berkeley, CA 94720}
\affil{$^2$ Lawrence Berkeley National Laboratory, 1 Cyclotron Road, Mail Stop 50-208, Berkeley, CA 94720}
\affil{$^3$ Hubble Fellow}
\affil{$^4$ UCO/Lick Observatory, Department of Astronomy and Astrophysics, University of California, Santa Cruz, CA 95064}
\affil{$^5$ Department of Physics, University of California, Berkeley, CA 94720}

\begin{abstract}
\noindent
\rightskip=0pt
We investigate the emission-line properties of galaxies with red
rest-frame colors (compared to the $g-r$ color bimodality) using
spectra from Data Release 4 of the Sloan Digital Sky Survey (SDSS). 
Emission lines are detected in more than half of the red galaxies. 
We focus on the relationship between two emission lines commonly used
as star formation rate 
indicators: \hal\ and \oiiw. Since \oii\ is the principal proxy for 
\hal\ at $z\sim1$, the correlation between them is critical for 
comparison between low-z and high-z galaxy
surveys.  We find a strong bimodality in \oii/\hal\ ratio
in the SDSS sample, which closely corresponds to the bimodality in
rest-frame color. 
Based on standard line ratio diagnostics, most (nearly all of the) 
line-emitting red galaxies have line ratios typical of various types of
Active Galactic Nuclei (AGN) --- most commonly ``low-ionization nuclear 
emission-line regions'' (LINERs), a small fraction of ``transition 
objects'' (TOs) and, more rarely, Seyferts. Only $\sim6\%$ of red galaxies have line ratios resembling star-forming galaxies. 
A straight line in the \oii-\hal\ EW diagram separates LINER-like galaxies from other categories, provides an effective classification tool complementary to standard line ratio diagnostics.
Quiescent galaxies with no detectable emission lines and those galaxies with LINER-like line ratios combine to form a single, tight red sequence in
color-magnitude-concentration space.  Other than modest differences in the luminosity range they span, these two classes are only distinguished from each other by line strength.
We also find that \oii\ equivalent widths in LINER- and AGN-like galaxies 
can be as large as that in star-forming galaxies. Thus, unless objects
with AGN/LINER-like line ratios are excluded, \oii\ emission cannot be 
used directly as a proxy for star formation rate; this is a particular 
issue for red galaxies. Lack of \oii\ emission is generally used to 
indicate lack of star
formation when post-starburst galaxies are selected at high redshift.
Our results imply, however, that these samples have been cut on
AGN properties as well as star formation, and therefore may
provide seriously incomplete sets of post-starburst galaxies.
Furthermore, post-starburst galaxies identifed in SDSS by requiring minimal 
\hal\ equivalent width generally exhibit weak but nonzero line emission 
with ratios typical of AGNs; few of them show residual star formation. 
This suggests that most post-starburst galaxies may harbor AGNs/LINERs.
\end{abstract}


\section{Introduction}
\oiiw\ emission line is a widely used, empirical star formation rate (SFR) 
indicator, especially at high redshift when \hal\ moves out of the optical 
window 
\citep[e.g.,][]{GallagherHB89,Kennicutt92,CowieSH96,CowieHS97,Ellis97,Hammer97,Hogg98,RosaGonzalezTT02,Hippelein03}. 
Although its luminosity is not directly coupled to the ionizing flux, and it 
is very sensitive to reddening and metallicity effects, it still can be 
empirically or theoretically calibrated through comparison with \hal. A 
variety of calibrations have been developed 
\citep{Kennicutt92,Kennicutt98, Kewley04, Mouhcine05, MoustakasKT05}, 
enabling SFR estimations 
from \oii\ to a reasonable accuracy for star-forming galaxies.

However, many red, elliptical galaxies at $z\sim0$ also have 
significant \oii\ and other line emission in their spectra 
\citep{Caldwell84, Phillips86}.  Does \oii\ also
indicate star formation in these galaxies? It has long been realized 
that star formation is not the only possible source of \oii\ emission in 
galaxies. Active Galactic Nuclei (AGN - especially LINERs), fast shockwaves, 
post-AGB stars, and cooling flows might also produce \oii\ emission. 
Thus, before we use \oii\ as a universal star formation indicator, we need
to know how often and to what degree \oii\ emission is contaminated by sources 
other than star formation, especially in red, elliptical galaxies.

The assumption that \oii\ measures star formation has been fundamental to 
many studies. For instance, 
the lack of \oii\ is often used as a criterion for post-starburst 
galaxy identification. Post-starburst galaxies, as their name suggests, 
are galaxies that underwent a strong recent star formation epoch
but have stopped forming stars. Their spectra can be modeled by a  
combination of an old stellar population (similar to that of a K giant star or 
an early-type galaxy) and a 
young population which is dominated by A stars \citep{DresslerG83}. 
Therefore, these galaxies are commonly
known as `K+A' or `E+A' galaxies. With time, these galaxies will 
display an early-type galaxy spectrum after all the A stars die out in 
1 Gyr. Thus, the 
study of these galaxies is important for understanding galaxy 
evolution, given the current uncertainty in how elliptical galaxies form. 

The identification of these post-starburst galaxies requires a total lack of 
emission lines, to be certain that star formation has truly stopped. 
Most works on post-starbursts
\citep[e.g.,][]{DresslerG83,Zabludoff96,Poggianti99,Balogh99,TranFI03,YangZZ04,
BlakePC04,TranFI04} 
have used \oii\ as the marker, either because it is the only emission line 
diagnostic available at high redshifts or for the sake of facilitating 
comparisons between low 
and high redshifts and among different authors.
Recently, several groups \citep{Goto03, Quintero04, Balogh05} have
employed \hal\ emission as the marker for star formation, rather than \oii, for 
samples of galaxies from the Sloan Digital Sky Survey (SDSS). 
Usually, a low equivalent width (EW) threshold on \oii, 
e.g., 2.5\AA, is used as the criterion for non-detection of such emission 
lines, according to the sensitivity in each survey. This reflects an 
underlying assumption that any emission line would be coming from 
starforming regions. If this assumption fails to hold---for instance, if 
the contribution of \oii\ from AGN is significant---the post-starburst sample 
defined by such a method would be incomplete because AGN would be 
mistakenly discarded as star-forming galaxies. Thus, using \oii\ in 
post-starburst galaxy studies requires a better understanding of the 
many origins of \oii\ in all types of galaxies.

In many studies of AGNs, the opposite question is asked: what fraction of \oii\ emission in AGN spectra comes from star formation? A variety of studies of high-ionization AGN (QSOs and Seyferts) have attempted to separate the contributions from AGN and star formation to line emission. It is of special interest to the understanding of the coevolution of the nuclear black holes and the bulges. \cite{Croom02} found by measuring the Baldwin effects \citep{Baldwin77} that most \oii\ emission in QSOs might be from the host galaxies instead of the AGN. \cite{Richards03} found that dust-reddened QSOs have stronger \oii\ emission than normal QSOs. One of the many possible explanations is that the host galaxies of those QSOs with higher extinctions have higher star formation rates. In contrast, studies by \cite{Ho05} show that in quasar spectra there is very little \oii\ emission beyond that expected from the AGN itself, indicating a supressed star formation efficiency in quasar host galaxies. 

Unlike the controversy in Type I QSOs, the picture is a little clearer for Seyfert 2s and Type II QSOs. \cite{Gu06} found that \oiii/\hb\ and \oiii/\oii\ ratios are lower in those Seyfert 2 galaxies with significant star formation, which indicates that star formation could contribute substantially or even dominate the \hb\ and \oii\ emission. Consistently, \cite{KimHI06} found that Type II QSOs exhibit significantly enhanced \oii/\oiii\ ratios relative to Type I QSOs. Combined with their high \oiii\ luminosities, the high \oii/\oiii\ ratio is best explained by a high level of star formation. 
However, the problem is far from settled. Although the \oii\ emission in QSO spectra could be a combination of two or many origins, we still have no idea of their proportions. In addition, QSO hosts are rare among galaxies. Narrow-line, low-luminosity AGN and LINERs (which may or may not be AGN) are much more abundant. The dominant origin of \oii\ emission among these galaxies is still unknown.

A variety of approaches can help to distinguish emission lines having 
different origins. 
One commonly used method, which is relatively reliable, is emission-line diagnostics \citep[etc.]{BPT,
VeilleuxO87, RolaTT97, KauffmannHT03}, as also used by many authors mentioned above in QSO studies.
This method is effective in distinguishing the
two major origins of emission: star formation versus AGN. We will focus on 
using this method to investigate the origin of \oii\ emission in red galaxies 
in this paper.

The large SDSS redshift survey provides a perfect sample for 
such a study. Its wide wavelength coverage covers many emission lines 
from \oii\ to \hal, from which many line ratio diagnostics can be created. 
At the same time, moderate resolution still allows reasonable line profile 
fitting to be conducted, giving good control of errors in the line 
strength measurement. Finally, its unprecedented huge sample size produces 
reliable statistics. Previous studies of line ratio diagnostics
\citep[e.g.,][]{RolaTT97} used much smaller datasets with less uniform sampling.

In this paper, we present the discovery of a bimodality in \oii/\hal\ ratio among galaxies. One mode is largely associated with star-forming galaxies. Galaxies in the other mode generally have line ratios similar to LINERs, with \oii\ emission produced by a mechanism not associated with ordinary star formation.
Narrow-line Seyferts and Transition Objects mostly fall in between the two dominant populations, consistent with the picture that both star formation and AGN (or some other source) might contribute substantially to the emission in these objects. The \oii/\hal\ bimodality we find also has important implications for post-starburst studies. 

The paper is structured as follows. In \S\ref{sec:data}, we describe the 
data used. (The details of our measurement methods are descibed in
the Appendices.) In \S\ref{sec:comp}, 
we compare \oii\ emission-line strength with \hal\ for all types of galaxies. 
We demonstrate that red and blue galaxies show different \oii-\hal\ 
correlations, which appear to reflect different emission origins.
Classification based on this \oii/\hal\ bimodality are made and investigated
later in \S\ref{sec:agn} with the standard
line ratio diagnostics. Possible origins of the emission in red galaxies 
are discussed. With the conclusions drawn from that, we show 
in \S\ref{sec:implications} how the selection of post-starbursts using \oii\ 
leads to an incomplete sample and explore the nature of emission in 
post-starbursts. We summarize in \S\ref{sec:summary}.
The cosmology used is a flat $\Lambda$ cold dark matter ($\Lambda$CDM) 
cosmology with density parameter $\Omega_m =0.3$.

\section{Overview of Data and Line Measurements}\label{sec:data} 
The SDSS \citep{York00, Stoughton02} is an ongoing imaging and 
spectroscopic survey that will eventually cover $\sim \pi$ steradians 
of the celestial sphere. It utilizes a dedicated 
2.5-m telescope at Apache Point Observatory. The imaging is done with 
five broadband filters in drift scan mode
\citep[$u,g,r,i, {\rm and}~z$;][]{Fukugita96,Stoughton02}.
Spectra are obtained with two fiber-fed spectrographs, covering the wavelength
range of 3800-9200\AA\ with a resolution of R $\sim 2000$. Fibers have a fixed
aperture of 3". 

The spectroscopic data used here have been reduced through the Princeton 
spectroscopic reduction pipeline (Schlegel et al.\ in prep), 
which produces the flux- and wavelength-calibrated 
spectra\footnote{http://spectro.princeton.edu/}. The redshift catalog of 
galaxies used is from the NYU Value Added Galaxy Catalog 
\citep[http://wassup.physics.nyu.edu/vagc/]{BlantonSS05}. 
K-corrections were derived using \cite{BlantonBC03}'s {\it kcorrect} 
code v3\_2. All the magnitudes used in this paper are K-corrected to a redshift
of $z=0.1$ by shifting the SDSS bandpasses to bluer wavelengths by a factor of
1.1 \citep{BlantonHB03}. Thus all magnitudes and color notations carry a 
superscript or subscript of 0.1 .

The results in this paper are based on the spectra of $\sim 400,000$ galaxies 
contained in the SDSS Data Release Four \citep[DR4]{SDSSDR4}.  
We employ a catalog of objects targeted by the main galaxy survey that 
have been spectroscopically confirmed as galaxies; QSOs are not included 
in this sample. Target selection for the main galaxy sample is described 
in \citet{Strauss02}.
The magnitude limit is $r=17.77$ in Petrosian magnitudes after correction
for the foreground galactic extinction following \cite{SchlegelFD98}.

For many plots (Figs.~\ref{fig:o2haew}, \ref{fig:o2haflux}, \ref{fig:sersic}, \ref{fig:oii_dist}, \ref{fig:ha_ak}, and \ref{fig:akgt02}) in this paper, we used a volume-limited 
sample of about 55,000 galaxies with $0.07<z<0.10$ and $M_{^{0.1}r}-5\log h$ 
brighter than -19.5, and to have well-measured spectra in the vicinity of \oii\ and \hal\ (using criteria described in the Appendix~\ref{sec:line}). However, 
in color-magnitude diagrams 
and line ratio diagnostic plots (Figs.~\ref{fig:cmd}, \ref{fig:redseq}, 
\ref{fig:bptred}, 
\ref{fig:o1ha_o3hb}, \ref{fig:o2o3_n2ha}, \ref{fig:o2hb_3f}, and \ref{fig:ak_bpt}), a magnitude limited sample with 
$0.05<z<0.10$ and $r$ brighter than 17.77 is used to cover a wider range 
in galaxy luminosity and metallicity.
The redshift ranges of these samples are chosen for several reasons.
First, selecting very nearby galaxies is unfavorable because Sloan spectra 
are taken with 
a 3" fiber which only covers a small fraction of each galaxy at low redshifts. 
Secondly, objects at redshifts greater than $z=0.15$ are disfavored because 
blue galaxies drop out 
the $r$-band selected sample quickly with increasing $z$. Additionally, the bad 
column on the CCD around 7300\AA\ heavily hinders the accurate measurements of
\hal\ at $z\sim0.114$ and other lines at higher $z$. The median S/N of the 
spectra in the volume-limited sample is 18.5 per pixel. 

For each galaxy, we subtract the stellar continuum and measure the emission 
line fluxes and equivalent widths for \oiiw, \hb, \oiiibw, \oiw, \hal, and 
\niibw. Zero point errors in the equivalent widths are removed and the corrections are propagated into fluxes. Reliable emission line detections are defined as 3 or more $\sigma$ detection in EW. For the technical details of our stellar continuum subtraction procedure, emission line measurements, and zero point determinations, see Appendix 
\ref{sec:contsub} and \ref{sec:line}.

\section{\oiiw\ vs. \hal}\label{sec:comp} 
\subsection{Color Bimodality}
\begin{figure}
\begin{center}
\includegraphics[totalheight=0.35\textheight]{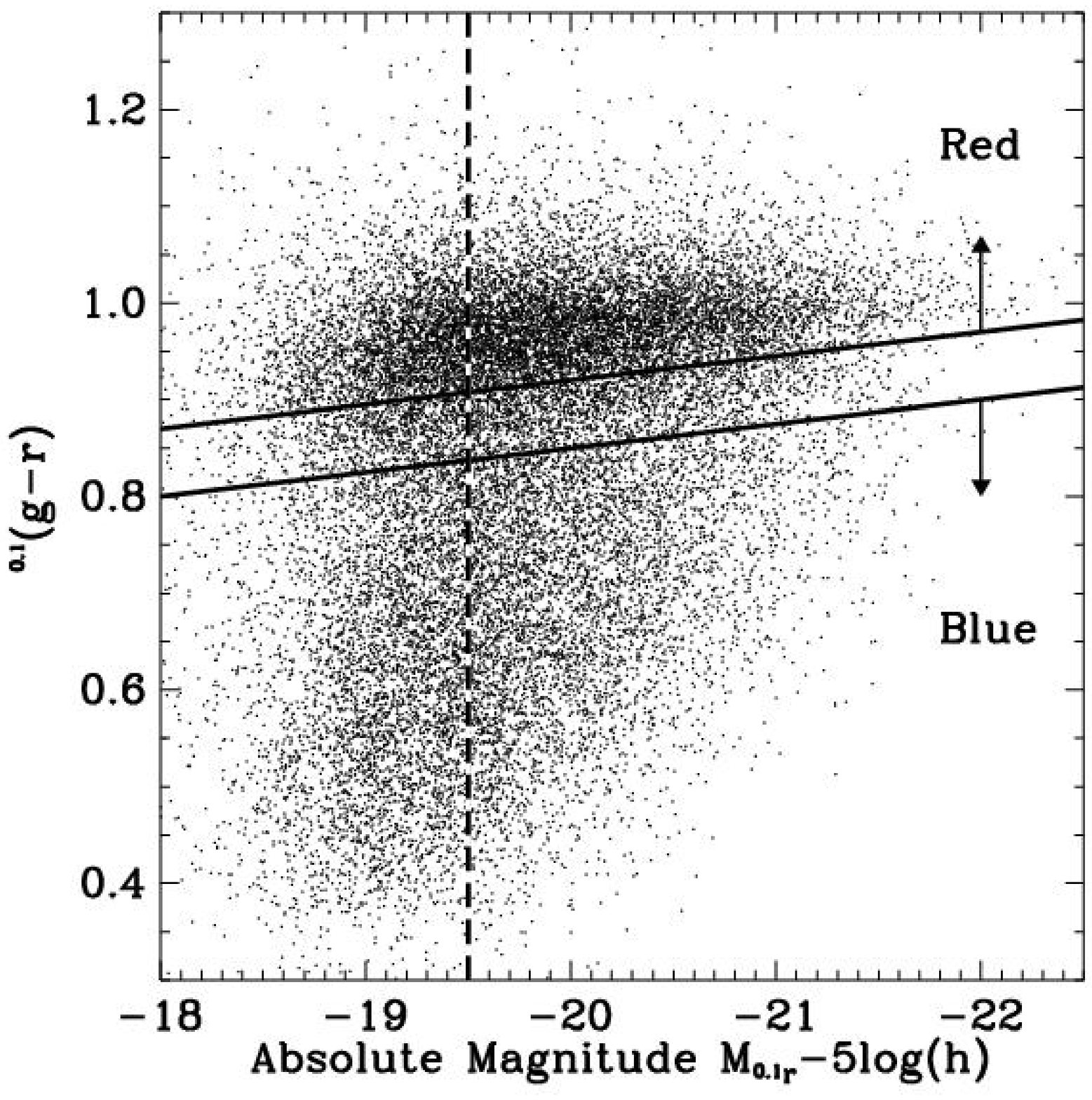}
\caption{Rest-frame color-magnitude diagram for all SDSS galaxies with redshifts between 
0.05
and 0.10. The two solid lines indicate the cut we adopt for separating red
and blue galaxies. Red galaxies are defined as the points above the upper line,
while blue galaxies are below the lower line. We leave a gap in between to
eliminate galaxies with potentially ambiguous status. The vertical dashed line
indicates the absolute magnitude limit for the volume-limited sample.}
\label{fig:cmd}
\end{center}
\end{figure}

Before diving into line comparisons, we recall that the color 
distribution of galaxies is highly bimodal, as have been shown repeatedly 
\citep[e.g.,][]{Strateva01,HoggBS02,BlantonHB03,BaldryGB04, BellWM04, WeinerPF05, 
Willmer05, Giallongo05}.
Fig.~\ref{fig:cmd} shows the color-magnitude diagram for all SDSS galaxies 
within 
$0.05<z<0.10$. We use this sample rather than the volume-limited sample simply 
to display a wider range in galaxy luminosity. 

In color-magnitude space, 
galaxies fall mainly into two categories. One category is 
the tight sequence between color and magnitude at the red end (larger value 
of $^{0.1}(g-r)$), commonly 
called the ``red sequence''; the other is the swath of points at bluer  
colors, which is sometimes referred to as the ``blue cloud''. 
Empirically, blue galaxies mostly have disk-dominated morphologies and are 
actively forming stars; while red galaxies mostly have bulge-dominated 
morphologies and are relatively quiescient in terms of star formation. 

Although the bimodality is obvious, it is impossible to entirely separate
the two populations by color. One possibility is that color provides a 
poor measurement of some more fundamental property that cleanly distinguishes
galaxies into two distinct populations. Alternatively, the absence of a clean 
separation in color may simply reflect the smooth 
evolution of one population into the other. It is also likely that both 
possibilities are involved. For now, we use conservative color cuts 
to define the samples of red and blue galaxies, as shown with the two solid 
lines in Fig.~\ref{fig:cmd} and desribed by the following inequalities.  We 
leave a gap in color between the two samples to reduce the importance of
color errors and galaxies with ambiguous status. 
We define these color cuts by:
\begin{eqnarray}
{\rm Red:~~} ^{0.1}(g-r) &>& -0.025 (M_{^{0.1}r}-5\log h) +0.42  \\
{\rm Blue:~} ^{0.1}(g-r) &<& -0.025 (M_{^{0.1}r}-5\log h) +0.35 .
\end{eqnarray}

\clearpage
\subsection{\oii/\hal\ Bimodality} \label{sec:bimodality}
In this paper, we will use both equivalent width (EW) and flux/luminosity 
measurements of emission lines. For line intensity ratios and SFR 
measurements, the use of flux is necessary, but fluxes are much more sensitive
to dust extinction than equivalent widths\footnote{This 
is not necessarily true in cases when the emission line region and the stars 
dominating the continuum have significantly different dust obscuration.}.
Furthermore, EW will not be affected by imperfect flux calibration.
On the other hand, to interpret the EW ratio between two lines, the continuum 
shape (i.e., the color of the galaxy) needs to be taken into account. 
In addition, line luminosity and EW describe different sorts of physical quantities: line
luminosity describes the total amount of emission, while EW reflects the 
strength of emission compared to the luminosity of the galaxy. We will 
primarily employ EW in this paper because it is not subject to variations in 
galaxy size (in contrast, line luminosity tends to scale with the total 
luminosity of a galaxy).
It is worth noting that the correlations in EW are often tighter than 
those in luminosity, for all these reasons.

\begin{figure*}
\begin{center}
\includegraphics[totalheight=0.75\textheight]{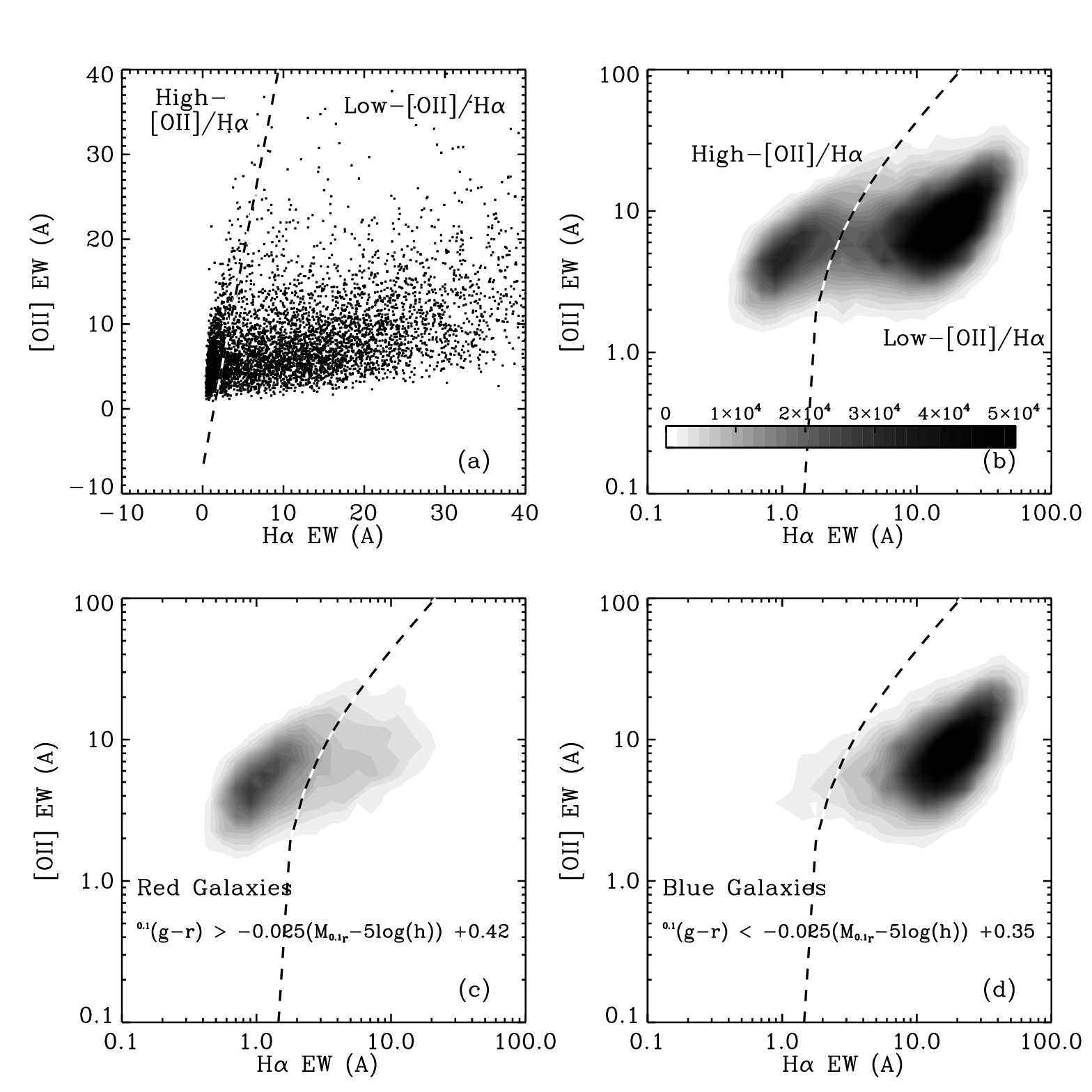}
\caption{{\bf (a)} The distribution of all SDSS galaxies in the 
volume-limited sample with $0.07<z<0.10$ in \hal\ and \oii\ equivalent 
width for those objects where both lines are detected. There are two \oii-\hal\ sequences with different slopes. 
The dashed line indicates an empirical cut (Eq.\ref{eqn:demar}) separating the 
two sequences and is reproduced in following panels. We refer to the population
to the left of the line as ``High-\oii/\hal\ galaxies'', and refer to the 
population to the right as ``Low-\oii/\hal\ galaxies''. {\bf (b)} Same as panel 
(a) but in logarithmic space. The shade indicates the number of points 
per sq. dex. The bimodality in \oii/\hal\ ratio is even more obvious in this panel.  
The dashed curve corresponds to the dashed line in panel (a).
{\bf (c)} Same as (b) but only for red 
galaxies, most of which are on the steep-slope sequence. Shade scale is the same as in panel (b). {\bf(d)} Same as (b) but only for blue galaxies, most of which are on 
the shallow-slope sequence. 
These generally have an \oii/\hal\ ratio consistent with an origin from star-forming HII 
regions (see \S\ref{sec:bimodality}). Dashed lines in all four panels indicate the same line, which we use throughout this paper to separate the two populations.
}
\label{fig:o2haew}
\end{center}
\end{figure*}

\begin{figure*}
\begin{center}
\includegraphics[angle=90,totalheight=0.33\textheight,viewport=0 0 330 730,clip]{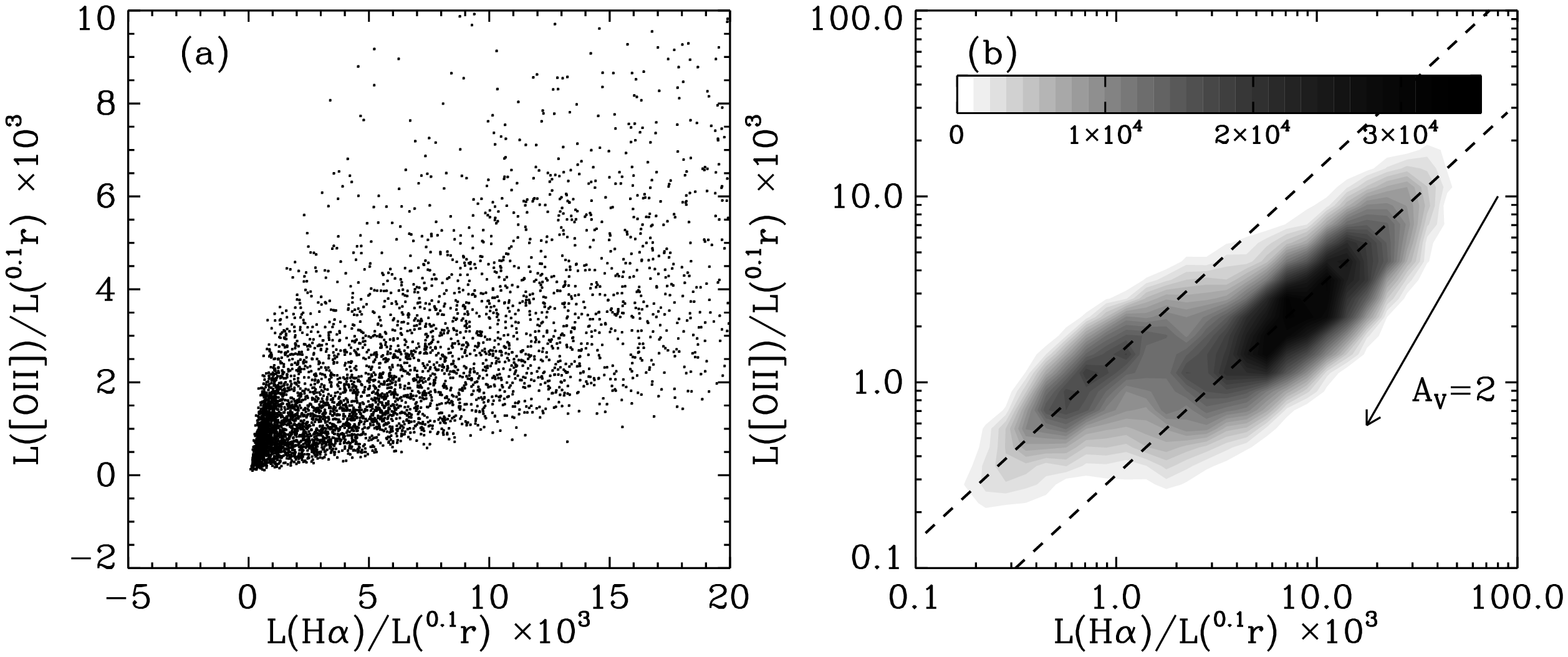}
\caption{{\bf(a)} [OII] line luminosity vs. \hal\ line luminosity for all 
galaxies 
in the volume-limited sample with $0.07 < z < 0.1$ detected in both lines. Line luminosities are 
divided by the SDSS $^{0.1}r$-
band luminosity to take out the size effect (see \S\ref{sec:bimodality}). The same bimodality 
exists as seen in EW plot. {\bf(b)} Same as panel (a) but in log-log space. 
The bimodality clearly stands out. The shade indicates the number of galaxies
per sq. dex. The two dashed lines show \oii/\hal\ ratios of 1.38 and 0.317, corresponding to the median ratio for each of the two modes, respectively.
The effect of $A_V=2$ of extinction (for $R_V=3.1$) is indicated by the arrow, as we do not correct for reddening.
}
\label{fig:o2haflux}
\end{center}
\end{figure*}

Figure \ref{fig:o2haew} compares [OII] EW with H$\alpha$ EW for galaxies in the volume-limited
sample ($0.07<z<0.10$, $M_{0.1r}-5\log h < -19.5$) which have both \oii\ and \hal\ positively detected. 
Clearly, there are two \oii-\hal\ sequences. One has a shallow slope 
and extends to very high \hal\ EW. 
The other sequence has a rather steep slope and very high concentration of 
points at low \hal\ EW, with \oii\ EW spanning a wide range.  

The bimodality in the \oii-\hal\ diagram becomes even more obvious if plotted 
on a
logarithmic scale, as shown in the upper right panel of Fig.~\ref{fig:o2haew}. 
Here the shading reflects the density of points in units of number per square 
dex. The \hal\ EW distribution can be well fit by two
log-normal distributions, one centered around 1\AA\ and the other centered 
around 14\AA. 
In linear space, the high-EW population spans a wide range of EWs, but it
becomes much more concentrated in log space.

The bottom two panels in Fig.~\ref{fig:o2haew} show \oii\ EW vs. \hal\ EW for 
the red galaxies and blue galaxies separately. The clear difference between 
the two populations shows immediately that most red 
galaxies reside on the steep-slope sequence, while most blue galaxies are on
the shallow-slope sequence. 
Clearly, the bimodality in \oii/\hal\ ratio echoes the color bimodality.
Since most blue galaxies are actively
forming stars, this is telling us that the shallow-slope sequence must have
an \oii/\hal\ ratio consistent with that produced in HII regions photoionized
by O and B stars in star-forming galaxies. 
What mechanism gives rise to the different ratio seen in red galaxies? 

Before addressing this question, we detour briefly to consider another issue.
Since we are thus far using equivalent width instead of line luminosity, 
one might imagine that the bimodality has little to do with emission 
strength; rather, differences in
eqivalent-width ratio might reflect only the different continuum ratios 
between stellar continua near \oii\ and 
\hal\ in red galaxies versus that of blue galaxies. Since red galaxies 
have higher continuum ratios of $F_c(\hal)/F_c(\oii)$,
they will have
a higher value of $\mbox{EW}(\oii)/\mbox{EW}(\hal)$ for the same emission line 
luminosity ratio.
This possibility can be checked by comparing L(\oii) to L(\hal) directly. 

As shown by \cite{Kennicutt89}, however, such a plot of luminosity versus 
luminosity can 
be quite misleading because of the size effect: for a given class of 
galaxies, we expect both L(\oii) and L(\hal) to each have a significant 
correlation with the stellar mass of the 
galaxies, especially at the low-mass end, which would veil the true 
relationship between \oii\ and \hal. 
This hidden variable can be easily removed by using specific line 
luminosity; here we divide the line luminosity by the Sloan $^{0.1}r$ band 
luminosity, which is a coarse proxy for the stellar mass. 
This effectively creates an EW, but one that does not include the continuum
color variation between red and blue galaxies.
As shown in Fig.~\ref{fig:o2haflux}(a), the same separation between the two 
sequences persists in a plot of L(\oii)/L($^{0.1}r$) vs. L(\hal)/L($^{0.1}r$), 
but with a smaller opening angle between them. The wider
opening angle in the equivalent width plot indeed reflects the different
continuum ratios between blue and red galaxies, but this is not the whole 
origin of the bimodality.
In panel (b) of Fig.~\ref{fig:o2haflux}, we plot L(\oii)/L($^{0.1}r$) 
vs. L(\hal)/L($^{0.1}r$) in 
logarithmic scale, analogous to the upper right panel in Fig.~\ref{fig:o2haew}.
The bimodality in \oii/\hal\ ratio again stands out clearly.
We omit the plots for subsamples separated in color, due to their close 
resemblance to the bottom panels of Fig.~\ref{fig:o2haew}.

Since the continuum ratios are not responsible for the \oii/\hal\ ratio 
bimodality, 
the question of the source of line emission in red galaxies persists. Can the 
difference between line ratios for blue and red galaxies in \oii/\hal\ ratio 
arise from reddening or metallicity dependences for star-forming regions? 
We investigated this 
possibility using the SFR(\hal) and SFR(\oii) calibrations derived by 
\cite{Kennicutt98} and \cite{Kewley04}, which explicitly take into account extinction and metallicity corrections:  
\begin{eqnarray}
{\rm SFR}(\hal) &=& 7.9\times10^{-42} L(\hal/{\rm erg s^{-1}}) M_\odot yr^{-1} 
\end{eqnarray}
and 
\begin{eqnarray}
{\rm SFR}(\oii) &=& {7.9\times10^{-42} L(\oii/{\rm erg s^{-1}}) \over
                        16.73-1.75[\log({\rm O/H})+12]} M_\odot yr^{-1}  .
\end{eqnarray}

By identifying SFR(\hal) with SFR(\oii), we get
\begin{equation}
L(\oii)/L(\hal) = 16.73-1.75[\log({\rm O/H})+12] .
\label{eqn:o2ha} 
\end{equation}

First, we can check that the \oii/\hal\ ratio seen among blue galaxies is consistent with that produced in star-forming HII regions. The galaxies to the right and below the demarcation in Fig.~\ref{fig:o2haew} have a median \oii/\hal\ flux ratio of 0.317, as indicated by the lower dashed line in Fig.~\ref{fig:o2haflux}b. Their median \hal/\hb\ ratio (the Balmer decrement) is 4.75. This corresponds to a median extinction $A_V$ of 1.60 (assuming $R_V = A_V/E(B-V) = 3.1$ and an intrinsic \hal/\hb\ ratio of 2.85 for case B recombination at $T=10^4{\rm K}$ and $n_e \sim 10^2-10^4 cm^{-3}$ \citep{Osterbrock89}). Applying the resulting extinction correction, we obtain a median intrinsic \oii/\hal\ ratio of 0.918. According to Eq.~\ref{eqn:o2ha}, this \oii/\hal\ ratio corresponds to a gas phase metallicity ($\log({\rm O/H})+12$) of 9.03, which is typical among star-forming galaxies in SDSS \citep{TremontiHK04}.

Next, we investigate the \oii/\hal\ ratios among the galaxies to the left and above the demarcation in Fig.~\ref{fig:o2haew}. These galaxies have a median \oii/\hal\ flux ratio of 1.38, as indicated by the upper dashed line in Fig.~\ref{fig:o2haflux}b. The median \hal/\hb\ ratio for these objects is 4.46, which corresponds to a median extinction in $A_V$ of 1.40 (following the same assumptions as for star-forming galaxies).
Applying the resulting extinction correction, we obtain a median intrinsic \oii/\hal\ ratio of 3.54. However, \cite{Kewley04} shows that the maximum \oii/\hal\ ratio which can be reached by star-forming regions is around 2.1, far below the ratio we find among red galaxies. If we were to take into account the metallicities inferred from the \nii/\oii\ ratio or (\oiiw+\oiiiw)/\hb\ ($R_{23}$) among the red galaxies, the discrepancy would become even larger.
As we will show in \S\ref{sec:agn} using full line ratio diagnostics, red galaxies in fact have line ratios characteristic of LINERs/AGNs, not star formation. Hints of this possible AGN origin of high \oii/\hal\ ratios can also be seen in earlier work by \cite{RolaTT97}, who suggested a new line diagnostic method using EW(\hb)and EW(\oii).

In Fig.~\ref{fig:o2haew}, a demarcation is shown separating the two \oii-\hal\ 
sequences; it is defined by 
\begin{equation}\label{eqn:demar}
{\rm EW}(\oii) = 5{\rm EW}(\hal) -7 .
\end{equation}
There is some overflow of points across the demarcation in the 
bottom panels of Fig.~\ref{fig:o2haew}. 
Roughly $3\%$ of blue galaxies in the volume-limited sample sit to the left 
of the demarcation. 
Nearly half of these
are post-starburst galaxies, which have had their star formation quenched to a 
fairly low level but still look blue because they contain 
relatively young stars. 
For the red galaxies, a much larger number of galaxies($\sim 30\%$) lie to the 
right of the line. As will be shown in \S\ref{sec:agn}, based on their other
line ratios, most of these galaxies are 
Transition Objects (TO), while a small fraction are Seyferts and dusty starforming galaxies. They will also be investigated further below. 

\subsection{Classification by \oii-\hal}\label{sec:subsample}

So far, we have been studying the galaxy distribution in \oii-\hal\ EW
plot for subsamples separated in their rest-frame colors.
While in selecting our blue and 
red subsamples we exclude a gap in $^{0.1}(g-r)$ color to reduce 
contamination, the overflow of points across the \oii-\hal\ demarcation is 
still significant for red galaxies. We can also investigate the bimodality 
from the other direction by dividing 
the sample according to position on the \oii-\hal\ EW plot and examine the
distribution of each subsample in color-magnitude space. 

\begin{figure*}
\begin{center}
\includegraphics[angle=90,totalheight=0.58\textheight]{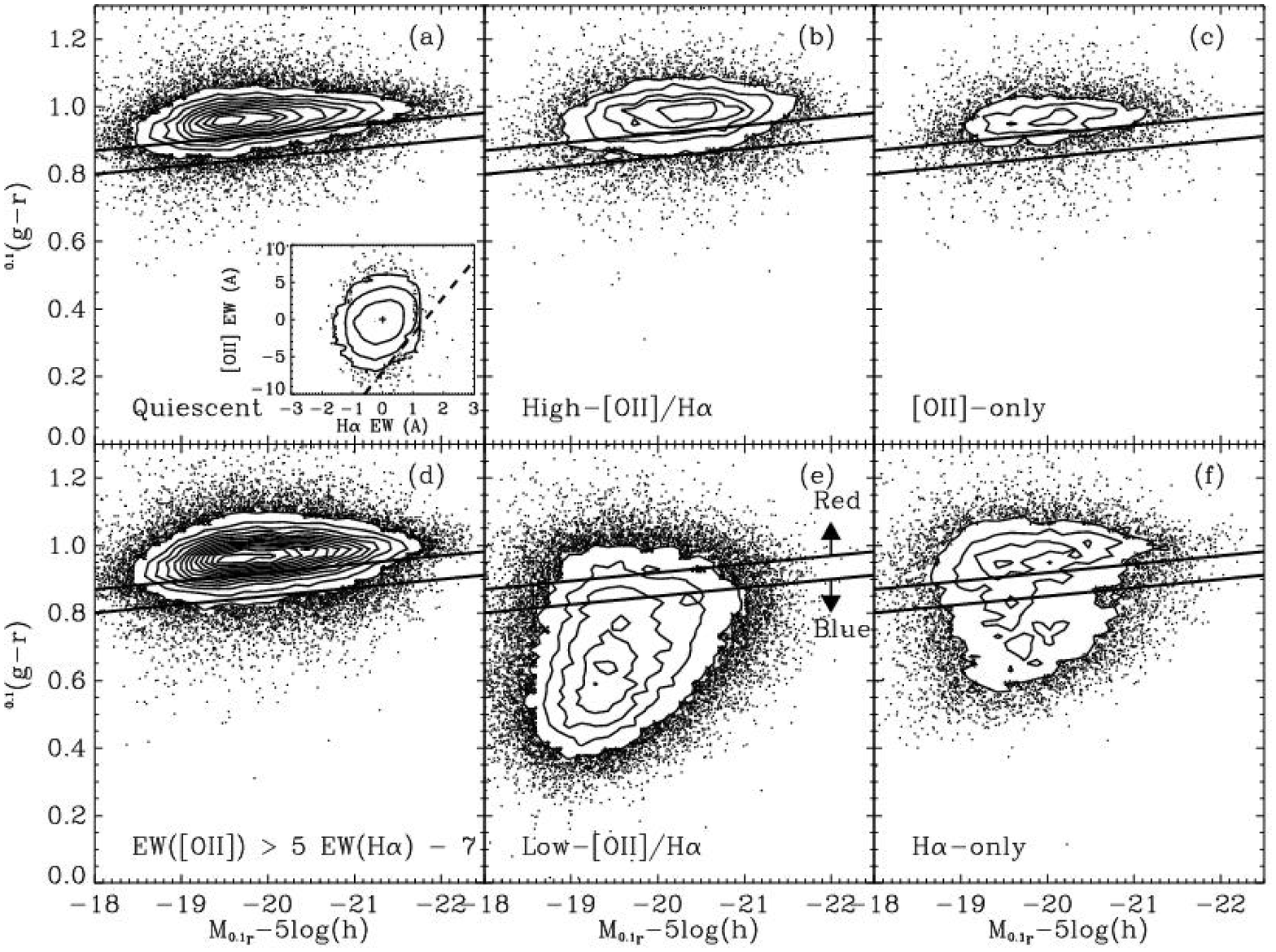}
\caption{
Color-magnitude diagrams for the quiescent galaxies (a), the high-\oii/\hal\ galaxies (b), the \oii-only galaxies (c), the galaxies to the left of the \oii-\hal\ demarcation (d), the low-\oii/\hal\ galaxies (e), and the \hal-only galaxies (f) in the magnitude limited sample ($0.05 < z < 0.10$). Quiescent galaxies, the high-\oii/\hal\ galaxies, and the \oii-only galaxies all follow a tight red sequence. Low-\oii/\hal\ galaxies cover not only the blue cloud but also part of the red sequence region. Contours are plotted at number densities of $1.6\times10^{4} n$ (halved in the panel b, c, and f) galaxies per mag$^2$, where $n$=1,2,..., starting from outermost contour. The inset in panel (a) shows the \oii-\hal\ EW distribution for the quiescent galaxies. They form a symmetric distribution around zero as expected. 
}
\label{fig:redseq}
\end{center}
\end{figure*}

We show the demarcation defined above in Eq.~\ref{eqn:demar} as the 
dashed line in Fig.~\ref{fig:o2haew}.
From now on, we refer to the population to the left of the line as 
``High-\oii/\hal\ galaxies'', and refer to the population to the right as 
``Low-\oii/\hal\ galaxies''. Note that there is a large population of galaxies which do not have both \oii\ and \hal\ detected and are not plotted here. Their EW measurements, if intrinsically zero, should have a symmetric Gaussian distribution around zero. These galaxies can be again classified into three groups based on the emission lines detectable in them:
\begin{enumerate}
\item 
Galaxies with \oii\ undetected and \hal\ detected. We label these ``\hal-only galaxies''. Note that other emission lines may be present in these objects' spectra besides \oii.  Similarly, the ``\oii-only'' galaxies defined below may have detectable lines besides \oii, but have no significant \hal\ emission.
\item
Galaxies with \oii\ detected and \hal\ undetected. These galaxies likely belong to the same population as the ``High-\oii/\hal'' galaxies and fall to the left of the demarcation line. We label these ``\oii-only galaxies''.
\item
Galaxies with neither \oii\ nor \hal\ detected. We label these galaxies as ``Quiescent Galaxies'' in the remainder of this paper. Their EW measurements form a symmetric distribution around (0,0) in the \oii-\hal\ diagram and fall to the left of our demarcation, as shown in the inset plot of Fig.~\ref{fig:redseq}. As illustrated in Fig.~\ref{fig:redseq}, quiescent galaxies have a color-magnitude distribution very similar to the ``High-\oii/\hal'' galaxies. 
\end{enumerate}

\begin{figure}
\begin{center}
\resizebox{3.5in}{!}{\includegraphics{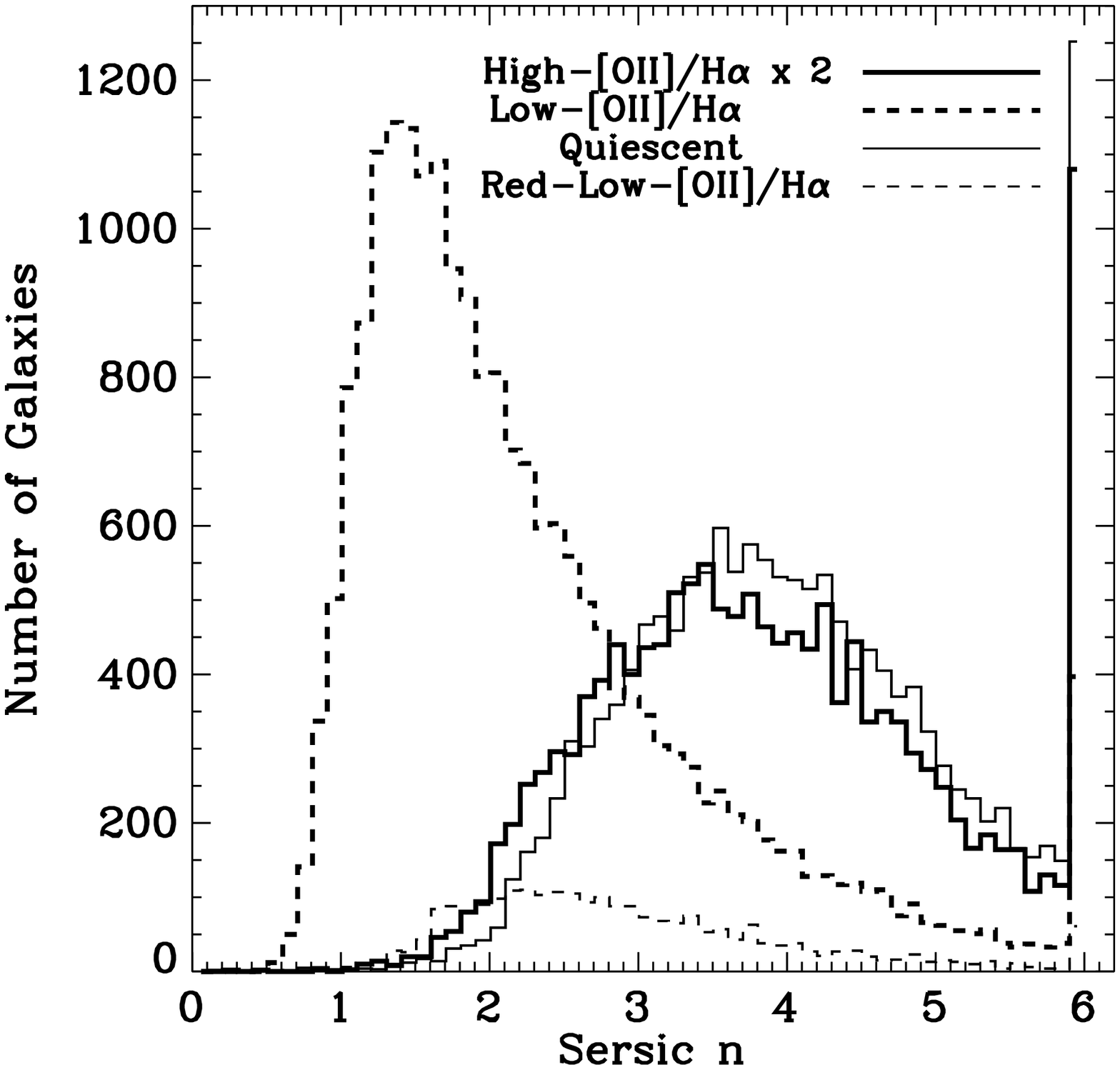}}
\caption{S\'{e}rsic index distributions of galaxies for four subsamples as defined in Table~\ref{tab:subsample}: high-\oii/\hal\ galaxies (thick-solid, height doubled for easy comparison), low-\oii/\hal\ galaxies (thick dashed), quiescent galaxies (thin-solid), and red-low-\oii/\hal\ galaxies (thin dashed). 
The high-\oii/\hal\ galaxies have a similar distribution to the quiescent galaxies, and very different from that of the low-\oii/\hal\ galaxies. The red-low-\oii/\hal\ galaxies have a distribution unlike either possible parent sample. 
}
\label{fig:sersic}
\end{center}
\end{figure}

Fig.~\ref{fig:redseq} shows the color-magnitude diagrams for these subsamples: quiescent galaxies (panel a), the high-\oii/\hal\ galaxies (b), \oii-only galaxies (c), the low-\oii/\hal\ galaxies (e), and \hal-only galaxies (f). Our color cuts are drawn here as the two straight lines for references. 
Both the quiescent galaxies and the high-\oii/\hal\ galaxies display a highly symmetric red sequence. The quiescent galaxies have a wider luminosity range and the high-\oii/\hal\ galaxies have a slight asymmetry in color, but they are otherwise nearly indistinguishable in color-magnitude space. As shown in Fig.~\ref{fig:sersic}, they also have similar S\'{e}rsic index
\footnote{S\'{e}rsic index is a seeing-corrected structural parameter describing the radial light profile of a galaxy. A S\'{e}rsic index of 4 corresponds to a de Vaucouleurs profile, and an index of 1 correponds to an exponential profile. The indices used here are derived by \cite{BlantonEH05}. One caveat is that these S\'{e}rsic indices are underestimates at high S\'{e}rsic index, e.g., n=3.5 for a de Vaucouleurs profile.} distributions. As expected, the \oii-only galaxies have similar color-magnitude distributions and S\'{e}rsic index distributions as the high-\oii/\hal\ galaxies. 
The combination of the three subsamples shown in Fig.~\ref{fig:redseq}a-c -- i.e., those galaxies to the left of the demarcation given by Eq.~\ref{eqn:demar} on \oii-\hal\ diagram -- together comprise a tight red sequence in the color-magnitude-concentration space, as shown in panel (d) of Fig.~\ref{fig:redseq}. 

The low \oii/\hal\ subsample, as expected, covers the blue cloud in the color-magnitude diagram (Fig.~\ref{fig:redseq}d). However, it also includes some galaxies with red-sequence-like colors, especially at the faint end. These red galaxies within the low \oii/\hal\ sample, which in fact are the overflow points in the lower left panel of 
Fig.~\ref{fig:o2haew}, also deserve special investigation. We will refer to 
them as ``Red-Low-\oii/\hal\ galaxies''. 
As shown in Fig.~\ref{fig:sersic}, these galaxies generally have lower
S\'{e}rsic indices than the overall red galaxy population. Their distribution does not resemble either of the possible parent samples; i.e., the red galaxy sample or the low-\oii/\hal\ sample. Thus, although these galaxies have red colors, they are likely to be a different population from the red sequence. The \hal-only galaxies (panel f) also extend to red colors. Without detection of \oii, their categorization is difficult; we will discuss them briefly later on. 
To recapitulate, we list in Table~\ref{tab:subsample} our sample definitions and the fraction of each category in the volume-limited sample. 

\begin{figure}
\begin{center}
\resizebox{3.5in}{!}{\includegraphics{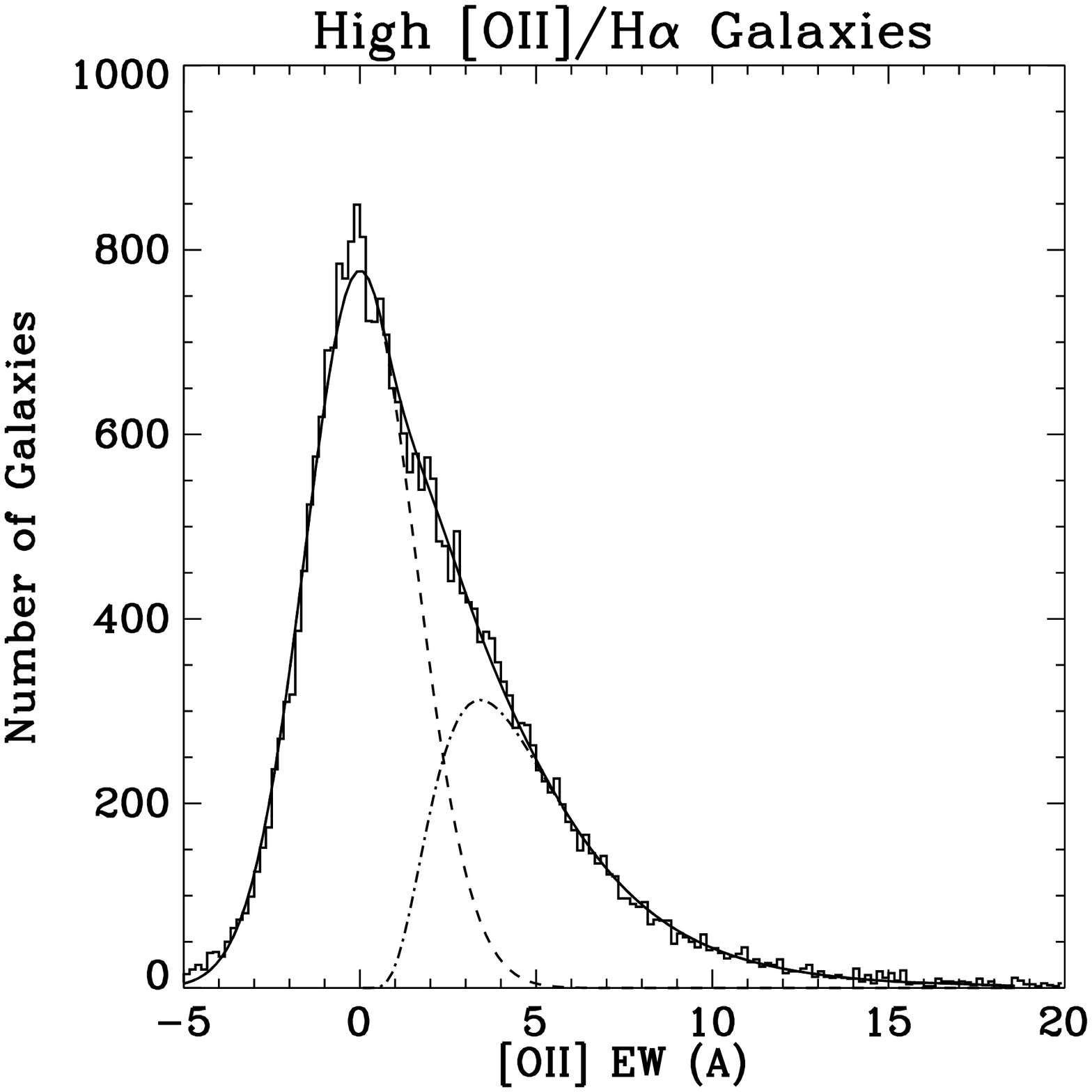}}
\caption{
The histogram shows the \oiiw\ EW distribution of all galaxies to the left of the demarcation (Eq.~\ref{eqn:demar}) in \oii-\hal\ diagram, including galaxies consistent with zero EW. It can be modeled well by the combination of a Gaussian distribution centered near zero (the dashed curve) and a log-normal distribution (the dot-dash curve). The sum of the two components is shown as the solid curve, which goes through the histogram.
The EW distributions of all other emission lines in this subset of galaxies can be modeled the same way, with similar results. As in the rest of this paper,plotted EWs are the values after zero point correction (see Appendix~\ref{sec:line} for details).
}  
\label{fig:oii_dist}
\end{center}
\end{figure}

The \oii\ EW distribution of all the galaxies to the left of the demarcation can be modeled well by the combination of a Gaussian distribution centered near zero and a log-normal distribution, as shown in Fig.~\ref{fig:oii_dist}. The two components correspond closely to the EW distributions of the quiescent population alone and the combined population of the high-\oii/\hal\ and the \oii-only galaxies, respectively; the fractional area covered by the log-normal distribution (35.4\%) is nearly identical to the fraction of galaxies with detectable \oii\ emission (35.5\%).

\begin{table*}[t]
\begin{center}
\begin{tabular}{|cr|c|c|c|c|}
\hline
\multicolumn{2}{|l|}{Name} & \multicolumn{2}{|c|}{Nonzero?} & Criteria & Percentage \\ \cline{3-4}
\multicolumn{2}{|l|}{ } & \oii & \hal & & \\ \hline
\multicolumn{2}{|l|}{High-\oii/\hal}  & Y & Y & $ {\rm EW}(\oii) > 5 {\rm EW}(\hal) - 7 $ & 13.3\% \\ \hline 
\multicolumn{2}{|l|}{Low-\oii/\hal}  &  Y & Y & $ {\rm EW}(\oii) < 5 {\rm EW}(\hal) - 7 $ & 38.0\% \\ 
 &Red-Low-\oii/\hal\  &    & & (as low-\oii/\hal, plus & ~~~(4.4\%)  \\
&   &  & & $^{0.1}(g-r) > -0.025(M_{^{0.1}r}-5\log h) +0.42 ) $ &  \\\hline
\multicolumn{2}{|l|}{\oii-only galaxies} & Y & N & & 5.2\% \\\hline
\multicolumn{2}{|l|}{\hal-only galaxies} & N & Y & & 15.7\% \\\hline
\multicolumn{2}{|l|}{Quiescent galaxies} &N & N & & 27.9\% \\ \hline
\end{tabular}
\caption{Subsample definitions and their fractions in the volume-limited sample; see \S\ref{sec:subsample} for details.
}
\label{tab:subsample}
\end{center}
\end{table*}

In the following section, we employ more line ratio diagnostics to 
investigate the source of the emission lines in red galaxies, specifically, 
the High-\oii/\hal\ population and the Red-Low-\oii/\hal\ population.

\section{Origin of [OII] emission in red galaxies} 
\subsection{Classifications by line ratio diagnostics} \label{sec:agn}
\begin{figure*}
\begin{center}
\includegraphics[angle=90,totalheight=0.33\textheight,viewport=0 15 320 750,clip]{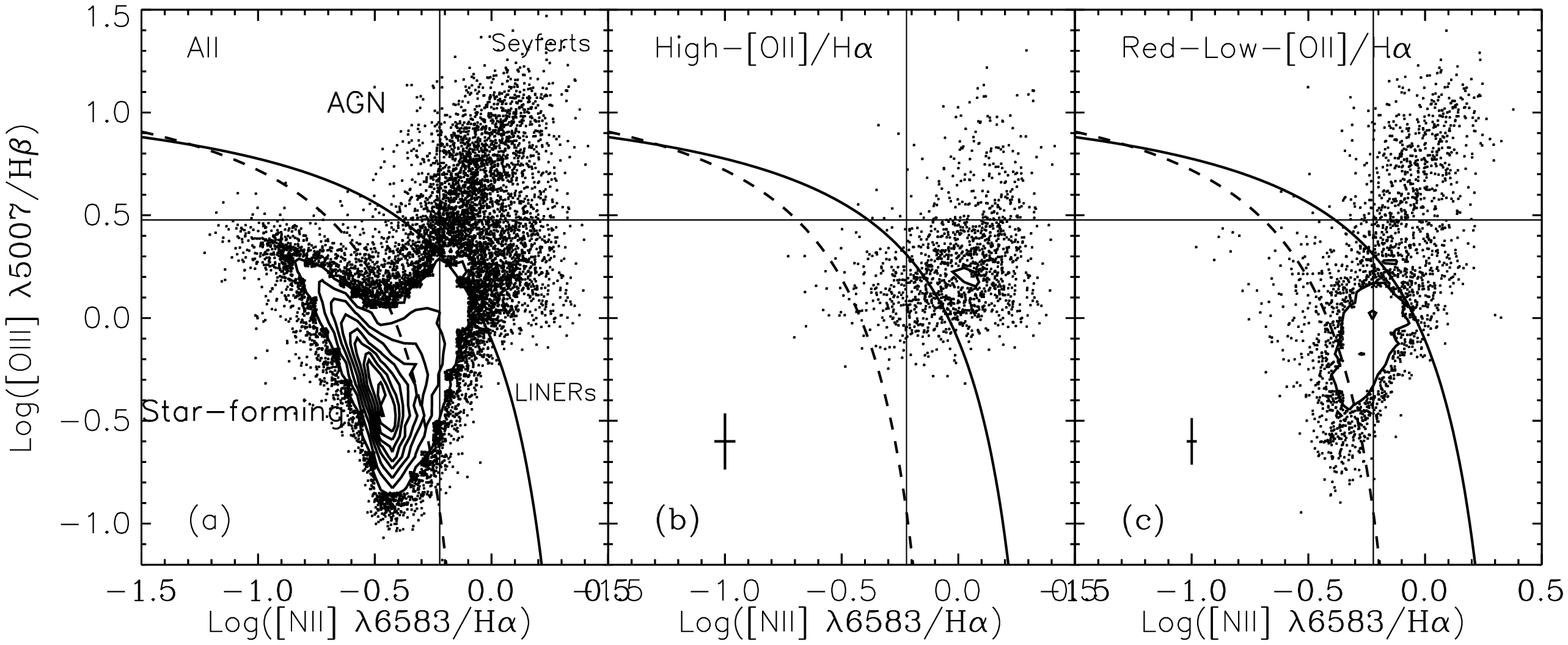}
\caption{{\bf (a)} Line ratio diagnostic diagram (BPT diagram) showing the 
emission-line flux ratio \oiii/\hb\ versus the ratio \nii/\hal\ for the 
magnitude-limited sample ($0.05< z<0.1$). This BPT digram, by construction, 
is very insensitive to the effects of extinction. All galaxies plotted 
have at least 2$\sigma$ detection on all four emission lines used. 
The solid curve shows the demarcation 
defined by \cite{KewleyDS01} to separate star-forming galaxies from AGN. 
The dashed curve shows the demarcation used by \cite{KauffmannHT03}. 
Conventional dividing lines for Seyferts and LINERs are indicated by the 
horizontal and vertical lines \citep{ShuderO81,VeilleuxO87}. 
Contours are plotted at number densities of $24,000n$ galaxies per unit area, where $n$=1,2,..., starting from the outermost contour.
{\bf (b)} Same diagram as panel (a) but only High-\oii/\hal\ galaxies 
are shown. These are centered in the LINER region. 
The $1\sigma$ measurement 
errors propagated to log-line-ratio is shown as the cross. 
Contours are plotted at number densities of $8,000n$ galaxies per unit area, where $n$=1,2,... .
{\bf (c)} Same diagram for the Red-Low-\oii/\hal\ sample. 
They lie to the left of the LINER region and extend down to the
star-forming region. 
Other than the Seyferts and the star-forming galaxies, most are Transition Objects, as shown in Fig.~\ref{fig:o1ha_o3hb}.
Contours indicate the same number density levels as in panel (b).
}
\label{fig:bptred}
\end{center}
\end{figure*}

Emission lines in galaxy spectra can be produced by a variety of mechanisms; 
Among them, photoionization by AGN and by hot stars in star-forming HII 
regions are two major mechanisms.
Baldwin, Phillips and Terlevich (\citeyear[hereafter BPT]{BPT}) pioneered the 
use of 
intensity ratios of two pairs of emission lines to distinguish starforming 
galaxies from AGNs. The method was further developed by \cite{VeilleuxO87}. 
The so-called BPT diagram of line ratios, such as the one shown in the panel 
(a) of Fig.~\ref{fig:bptred}, has become a standard tool for 
determining the source of line emission. 

All the galaxies plotted in Fig.~\ref{fig:bptred} have at least 2$\sigma$ 
detections of all emission lines considered (\hb, \oiiibw, \hal, \niibw).
Two sequences of emission-line galaxies dominate this diagram; one 
corresponds to
the star-forming galaxies, and the other to AGN. The star-forming sequence
is the curved concentration of points that extends from center bottom to upper 
left; its shape reflects the correlation between ionization parameter and  
metallicity in star-forming galaxies \citep{KewleyDS01}.
The concentration of points on the right part of the plot is identified with 
AGNs \citep[e.g.,][]{BPT,VeilleuxO87,KauffmannHT03}. In studies employing 
optical 
emission lines from large aperture spectroscopy, one usually cannot confirm 
that the emission lines are associated with a
central supermassive black hole. From now on, we define a ``true AGN'' as 
a system in which the radiation is ultimately powered by accretion 
onto a supermassive black hole. In the remainder of this paper,
we will refer to galaxies with line ratios characteristic of true AGN as
``AGN-like galaxies''.

The theoretical modeling of starburst galaxies can produce a boundary
for the locus of star-forming galaxies on the BPT diagram, which is called the 
extreme 
starburst classification line. The classification line given by 
~\cite{KewleyDS01} is indicated by the solid curve in Fig.~\ref{fig:bptred}.
As seen in this figure, that would serve as a very conservative cut for 
selecting AGN-like galaxies. An alternative empirical demarcation has been 
suggested by \cite{KauffmannHT03}, which is indicated by the dashed curve in 
Fig.~\ref{fig:bptred}. 

\begin{figure*}
\begin{center}
\includegraphics[angle=90,totalheight=0.33\textheight,viewport=0 15 320 750,clip]{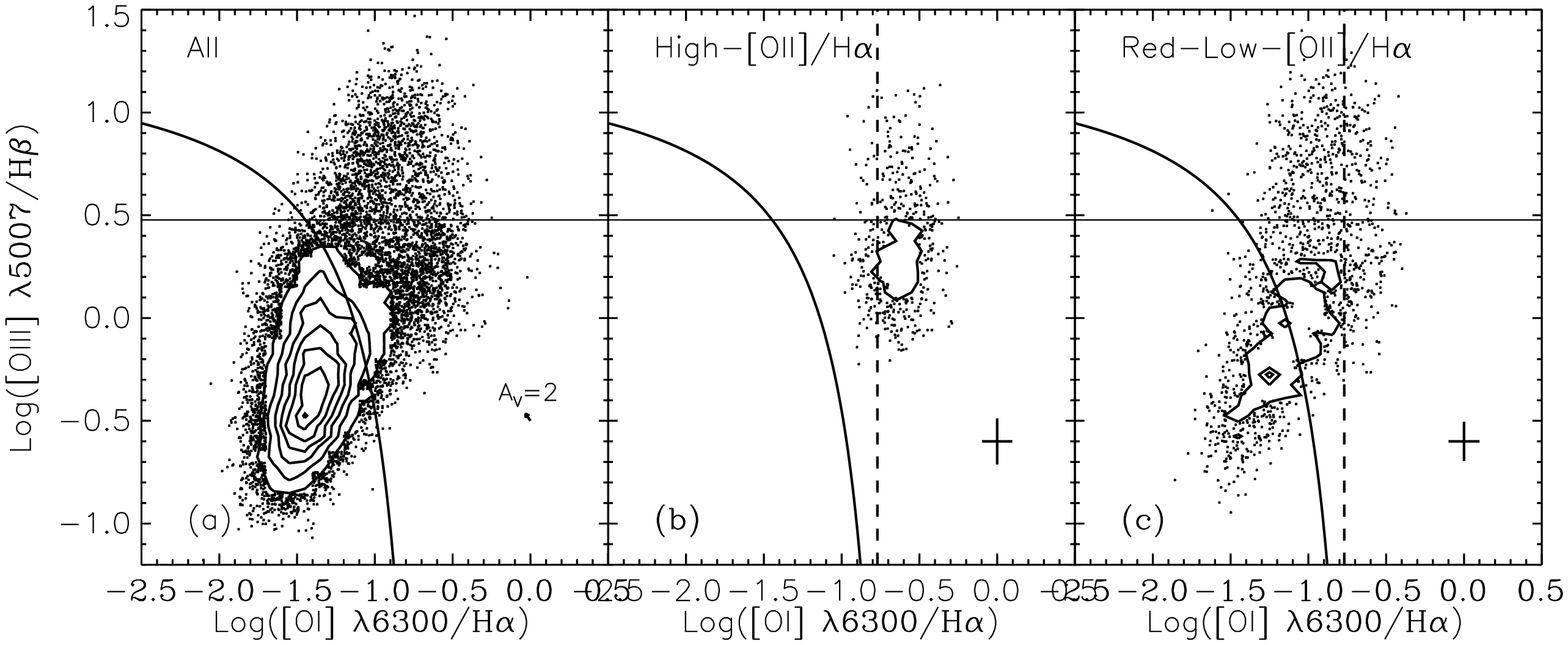}
\caption{{\bf (a)} BPT diagram showing the emission-line flux ratio 
\oiiibw/\hb\ versus
the ratio \oiw/\hal\ for the magnitude-limited sample ($0.05<z<0.1$). 
The effect of $A_V=2$ of extinction (for $R_V=3.1$) is indicated by the 
little arrow.
All galaxies plotted here have at least 2$\sigma$ 
detections on all four emission lines. The solid curve shows the demarcation 
defined by \cite{KewleyDS01} to separate star-forming galaxies (to the left and below) from AGN (to the right and above).  
The horizontal line is the dividing line 
between Seyferts (above) and LINER/TOs (below) as defined by \cite{ShuderO81} and \cite{VeilleuxO87}.
Contours indicate number density levels of $10,000n$ ($n$=1,2,...) galaxies per unit area.
{\bf (b)} The same diagram for the High-\oii/\hal\ sample. The vertical line is 
the dividing line between Transition Objects and LINERs given by 
\cite{HoFS97III}.
The High-\oii/\hal\ sample lies almost entirely in the LINER region, with 
scatter similar to the measurement errors.
The $1\sigma$ measurement 
errors propagated to log-line-ratio is shown as the cross. 
Contours indicate number density levels of $2,400n ~(n=1,2,...)$ galaxies per unit area.
{\bf (c)} The same diagram for the Red-Low-\oii/\hal\ sample. These objects 
lie to 
the left of the TO/LINER dividing line and thus are indeed Transition 
Objects, plus a few Seyferts and star-forming galaxies. 
Contours indicate the same number density levels as in panel (b).
}
\label{fig:o1ha_o3hb}
\end{center}
\end{figure*}

Traditionally, AGNs are further divided into several sub-categories according 
to their positions in the BPT diagram: Seyferts, Low-Ionization Nuclear 
Emission-line Regions (LINERs) and Transition Objects (TOs). 
Traditional definitions for Seyferts and LINERs are indicated by the
horizontal and vertical lines \citep{ShuderO81,VeilleuxO87} in 
Fig.~\ref{fig:bptred}.
Ever since the discovery of LINERs \citep{Heckman80}, it has remained uncertain
whether they are true AGN.  Recently, it has become clear
that at least some LINERs are indeed accretion-powered
systems \citep[and references therein]{Filippenko03, Ho04}. However, LINER-type 
line ratios are also found coming from extended regions in some early-type 
galaxies \citep{HeckmanBvB89,Goudfrooij94,Goudfrooij99}, in support 
of other ionization mechanisms, such as shockwaves, cooling flows, hot old stars, etc.
Transition Objects are galaxies that satisfy most of the LINER criteria but
have weaker \oiw/\hal\ ratio \citep{FilippenkoT92,HoFS93}. The dividing line between 
LINERs and TOs in \oi/\hal\ is indicated by the vertical dashed line in 
Fig.~\ref{fig:o1ha_o3hb}, which shows another useful BPT diagram 
with \oiiibw/\hb\ plotted against \oiw/\hal.
The nature of Transition Objects still remains unclear.  

Now we examine the position of the two classes of red galaxies in the BPT 
diagrams. We look 
at the High-\oii/\hal\ sample and the Red-Low-\oii/\hal\ sample separately
because they have different \oii-\hal\ correlations. In the two BPT diagrams
(Fig.~\ref{fig:bptred} \& \ref{fig:o1ha_o3hb}), the two samples are 
displayed by themselves in panels (b) and (c). In both BPT diagrams, 
the High-\oii/\hal\ sample (panel b) has a nearly symmetric distribution 
centered on the region classified as LINERs. The scatter is due to intrinsic
variations as well as measurement errors. The median errors are indicated 
on the plot.
Galaxies in the Red-Low-\oii/\hal\ sample 
mostly fall into the transition region in both BPT diagrams (panel (c) in 
each). Fig.~\ref{fig:o1ha_o3hb} shows clearly that most of these 
galaxies are in fact Transition Objects as defined by \cite{HoFS97III}, 
while a smaller fraction are Seyferts and star-forming galaxies. 

Comparing \oiiw\ with \oiw, one can see the similarity in their relations to
\hal: both \oii/\hal\ and \oi/\hal\ ratios provide a separation between LINERs and other categories. In fact, the bimodality in \oi/\hal\ is employed in one of the BPT 
diagrams to distinguish AGNs from star-forming HII regions.
Unfortunately, \oiw\ is often too weak to detect, which hinders 
the use of this particular BPT diagram for the full sample. As we have shown above,
the strong bimodality in EW(\oii)/EW(\hal) (or similarly, EW(\oii)/EW(\hb) ) 
might make it a good alternative, despite the wide separation of the two lines 
in wavelength. 
We propose the use of the demarcation in the EW(\oii)-EW(\hal) plane
shown in Fig.~\ref{fig:o2haew} to separate LINER-like galaxies 
from other line-emitting objects.

\begin{figure*}
\begin{center}
\includegraphics[angle=90,totalheight=0.36\textheight,viewport=0 20 350 735,clip]{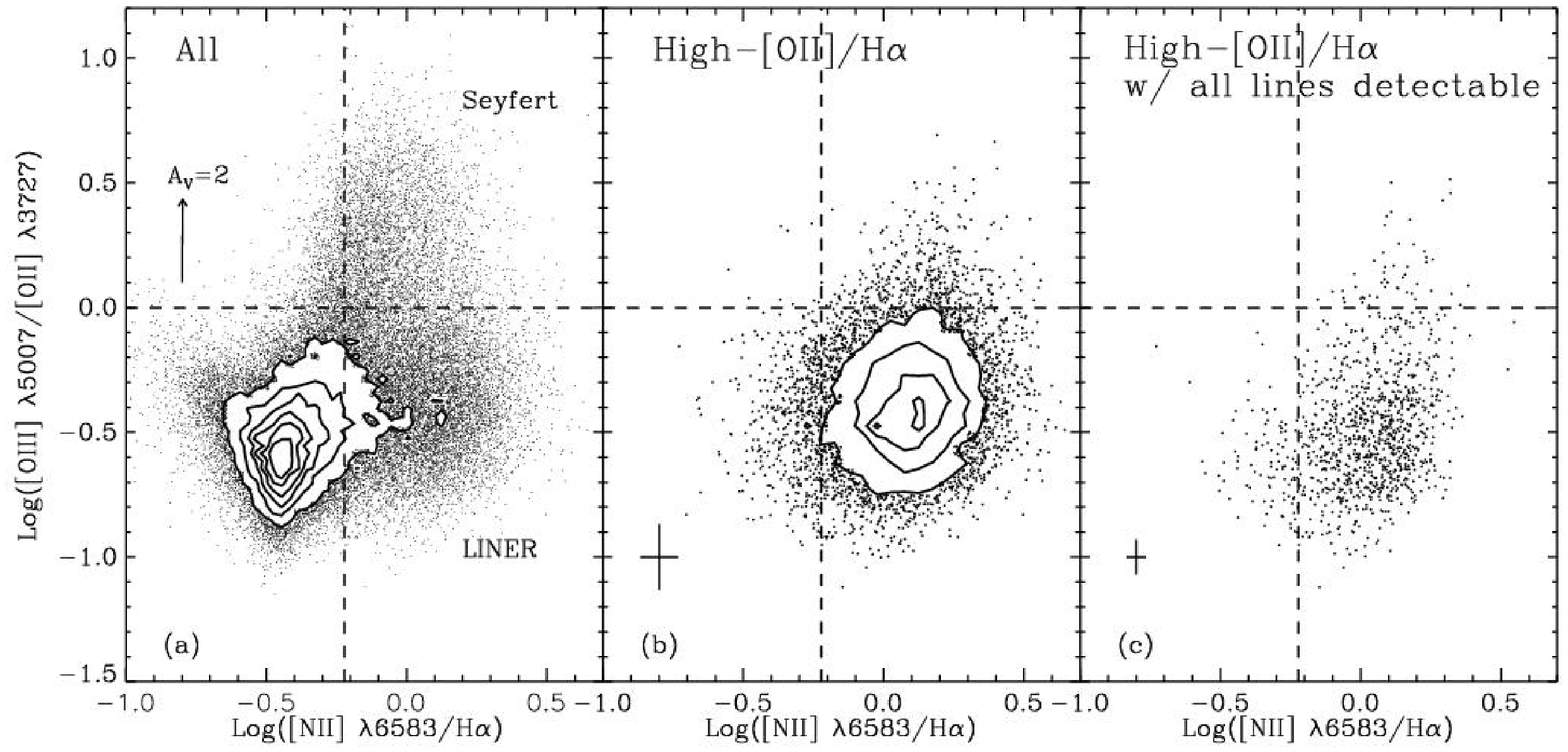}
\caption{{\bf (a)} Line ratio diagnostic diagram (BPT diagram) showing the 
emission-line flux ratio \oiii/\oii\ versus the ratio \nii/\hal\ for the 
magnitude-limited sample ($0.05< z<0.1$). All galaxies plotted 
have at least 2$\sigma$ detections on all four emission lines. 
Traditional definitions for Seyferts and LINERs are indicated by the 
horizontal and vertical lines \citep{Heckman80,ShuderO81,VeilleuxO87}. 
Contours are plotted at number densities of $45,000n$ galaxies per unit area, where $n$=1,2,.... .
The effect of $A_V=2$ of extinction (for $R_V=3.1$) is indicated by the arrow.
{\bf (b)} Same diagram as panel (a) but only High-\oii/\hal\ galaxies 
are shown. These are centered in the LINER region. 
The $1\sigma$ measurement errors propagated to log-line-ratio is shown as the cross. 
Contours are plotted at number densities of $25,000n$ galaxies per unit area, where $n$=1,2,... . {\bf (c)} Same diagram but only high-\oii/\hal\ galaxies which also have at least 2$\sigma$ detections on \oi\ and \hb\ are plotted. They have similar distributions as those without detectable \oi\ or \hb, only biased slightly against high \nii/\hal\ ratio and high \oiii/\oii\ ratio.
}
\label{fig:o2o3_n2ha}
\end{center}
\end{figure*}

Not all of the high-\oii/\hal\ galaxies are plotted on the two BPT diagrams(Fig.~\ref{fig:bptred} \& \ref{fig:o1ha_o3hb}). Because \oiw\ and \hb\ are much weaker than strong emission lines like \oiiw\ and \hal, requiring them to be detected significantly reduces the sample size, especially among the high-\oii/\hal\ population. Do such requirements bias our sample selection? We test this with a BPT diagram (Fig.~\ref{fig:o2o3_n2ha}) which only involves the strong lines: \oiii/\oii\ vs. \nii/\hal. Panel (a) of Fig.~\ref{fig:o2o3_n2ha} shows the distribution for all galaxies with at least 2-$\sigma$ detections on all of the four strong lines. Galaxies identified as star-forming from other BPT diagrams are again found
from bottom to middle left. The horizontal line indicates the \oiii/\oii\ division originally defined by \cite{Heckman80} to separate LINERs from Seyferts. The vertical line is a commonly adopted limit for separating LINERs and Seyferts from star-forming galaxies \citep[e.g.,][]{HoFS97III}. 
In panel (b) of Fig.~\ref{fig:o2o3_n2ha}, only high-\oii/\hal\ galaxies are plotted. It is worth noting that the high-\oii/\hal\ galaxies again fall into the LINER region. 
In panel (c), we only plot high-\oii/\hal\ galaxies which also have at least 2-$\sigma$ detections on \oi\ and \hb, i.e., those which are also plotted in Fig.~\ref{fig:bptred} \& \ref{fig:o1ha_o3hb}. These galaxies have very similar distributions as those without detectable \oi\ or \hb, only biased slightly against high \nii/\hal\ ratio and high \oiii/\oii\ ratio. The former is likely due to the difficulty of measuring \hb\ for galaxies with very low \hal\ and thus high \nii/\hal\ ratio. The latter is probably due to the difficulty of measuring \oi\ for galaxies with relatively high ionization. Thus, requiring \oi\ and \hb\ detections only biases the sample slightly, and in expected ways. We thus believe that nearly all high-\oii/\hal\ galaxies have LINER-like line ratios, regardless of their \oi\ or \hb\ detectability.

The high-\oii/\hal\ galaxies, as shown above, have relatively uniform line ratios in \oii/\hal, \oiii/\hb, \nii/\hal, \oi/\hal, and \oiii/\oii, which altogether identify them as LINER-like. As found in \S\ref{sec:subsample}, the high-\oii/\hal\ galaxies occupy the same region in the color-magnitude space as quiescent galaxies do and they share the same distribution in galactic concentration. 
Since they only differ in emission line strengths, we speculate that these LINER-like galaxies have a close evolutionary relationship with the quiescent red sequence galaxies. 
They might be the immediate progenitors of the quiescent red galaxies --- a scenario in which the LINER phase lasts for some period of time when the galaxy first arrives on the red sequence (Graves et al., in prep). Alternatively, it could be that LINER phases are intermittent throughout the life of a quiescent galaxy. As pointed out in \S\ref{sec:subsample}, the combination of the LINER-like (high-\oii/\hal) galaxies and the quiescent galaxies produces a uniform red sequence; the red sequence is effectively a LINER/Quiescent sequence. 

We turn now to the low-\oii/\hal\ galaxies. As was the case previously, not all of the low-\oii/\hal\ galaxies are plotted on the BPT diagrams because of S/N limitations. But as we found before, in the \oiii/\oii\ vs. \nii/\hal\ diagram, the distribution of red-low-\oii/\hal\ galaxies is also found to be independent of their \oi\ or \hb\ detectability. Their line ratios vary from star-forming, to Transition Objects, to Seyferts. To quantify the relative fraction of each category, we use \nii/\hal\ ratio, which is most commonly detectable among this population. As listed in Table~\ref{tab:red_breakdown}, a little more than 1/3 of these have $\nii/\hal < 0.6$, similar to dusty star-forming galaxies, while the rest have $\nii/\hal > 0.6$, resembling TO/Seyferts. 

The \oii-only galaxies likely belong to the high-\oii/\hal\ category intrinsically, as they occupy the same color-magnitude space, have the same S\'{e}rsic index distribution, and have lower limits on \oii/\hal\ ratio higher than typical high-\oii/\hal\ galaxies. Therefore, we suspect they also belong to the LINER-like category.

The case for the \hal-only galaxies is more complicated, since the ratio of the EWs at which \oii\ and \hal\ are detectable is similar to the \oii/\hal\ EW ratio of a typical high-\oii/\hal\ galaxy. The \hal-only population will include both high-\oii/\hal\ and low-\oii/\hal\ galaxies. Among them, we can only separate star-forming galaxies out using the \nii/\hal\ ratio, the only line ratio available in most of these galaxies; see Table~\ref{tab:red_breakdown}.

Previous studies of emission lines in early-type galaxies had already shown 
that \oiiw\ emission is detected in roughly half of all early-type 
galaxies \citep{Mayall58,Caldwell84, Phillips86}. Most of these
galaxies in which multiple emission lines are detected have LINER-like line 
ratios \citep{Phillips86, Goudfrooij94, Zeilinger96}.
With a much bigger sample, we find from SDSS data that $\sim38\%$ of all red galaxies
in the volume-limited sample have \oiiw\ positively detected. As shown above, 
the red-sequence population actually has two major sub-components: 
the LINER/Quiescent sequence and the TO/Seyferts. The former dominates the red galaxy population; the LINER-like population is one of the two main components in the \oii-\hal\ bimodality. Table~\ref{tab:red_breakdown} gives the detailed population breakdown of red galaxies in the volume-limited sample. Emission lines are detected in 52.2\% of all red galaxies; LINER-like galaxies comprise {\it at least} 28.8\%, dusty-starforming galaxies comprise about 6.5\%, while TOs and Seyferts are {\it less than} 16.8\%.

\begin{table*}[t]
\begin{center}
\begin{tabular}{|c|c|c|r|c|p{3in}|}
\hline
\multicolumn{2}{|c|}{Nonzero?}  & \oii/\hal\  & Number & Percentage & Category\\ \cline{1-2}
\oii & \hal & & & &  \\ \hline
N & N & & 12913 & 47.8\% & Quiescent \\ \hline
N & Y & Uncertain & 3886 & 14.4\% & LINERs, TOs and Seyferts (\nii/\hal~$>0.6$, 11.1\%); Dusty SF (\nii/\hal~$<0.6$, 3.2\%) \\ \hline
Y & N & High& 2209 & 8.2\% & Possibly LINER-like \\ \hline
Y & Y & High & 5571 & 20.6\%  & LINER-like \\ \cline{3-6}
 & & Low & 2421 & 9.0\% & TOs and Seyferts (\nii/\hal~$>0.6$, 5.7\%); Dusty SF (\nii/\hal~$<0.6$, 3.3\%) \\ \hline
\multicolumn{2}{|c|}{Total} &  & 27000 & 100\% & All red galaxies\\ \hline
\end{tabular}
\caption{Detailed population breakdown of red galaxies in the volume limited sample, see \S\ref{sec:agn}. 
}
\label{tab:red_breakdown}
\end{center}
\end{table*}

\subsection{Possible Sources of the Emission in High-\oii/\hal\ Galaxies}\label{sec:agn_mechanisms}
As discussed in the introduction, many recent studies suggest that star formation could contribute substantially to the \oii\ and \hb\ emission in Type II QSOs and Seyfert 2 galaxies. Is this also the case for LINERs?  
We have shown in \S\ref{sec:bimodality} that the \oii-\hal\ ratio in the high-\oii/\hal\ population cannot be reached by star formation alone. Here, we provide additional evidence using a BPT-like diagram involving \oiiw\ to show that star formation does not contribute substantially to \oii\ emission in these galaxies. 

\begin{figure}
\begin{center}
\includegraphics[totalheight=0.35\textheight]{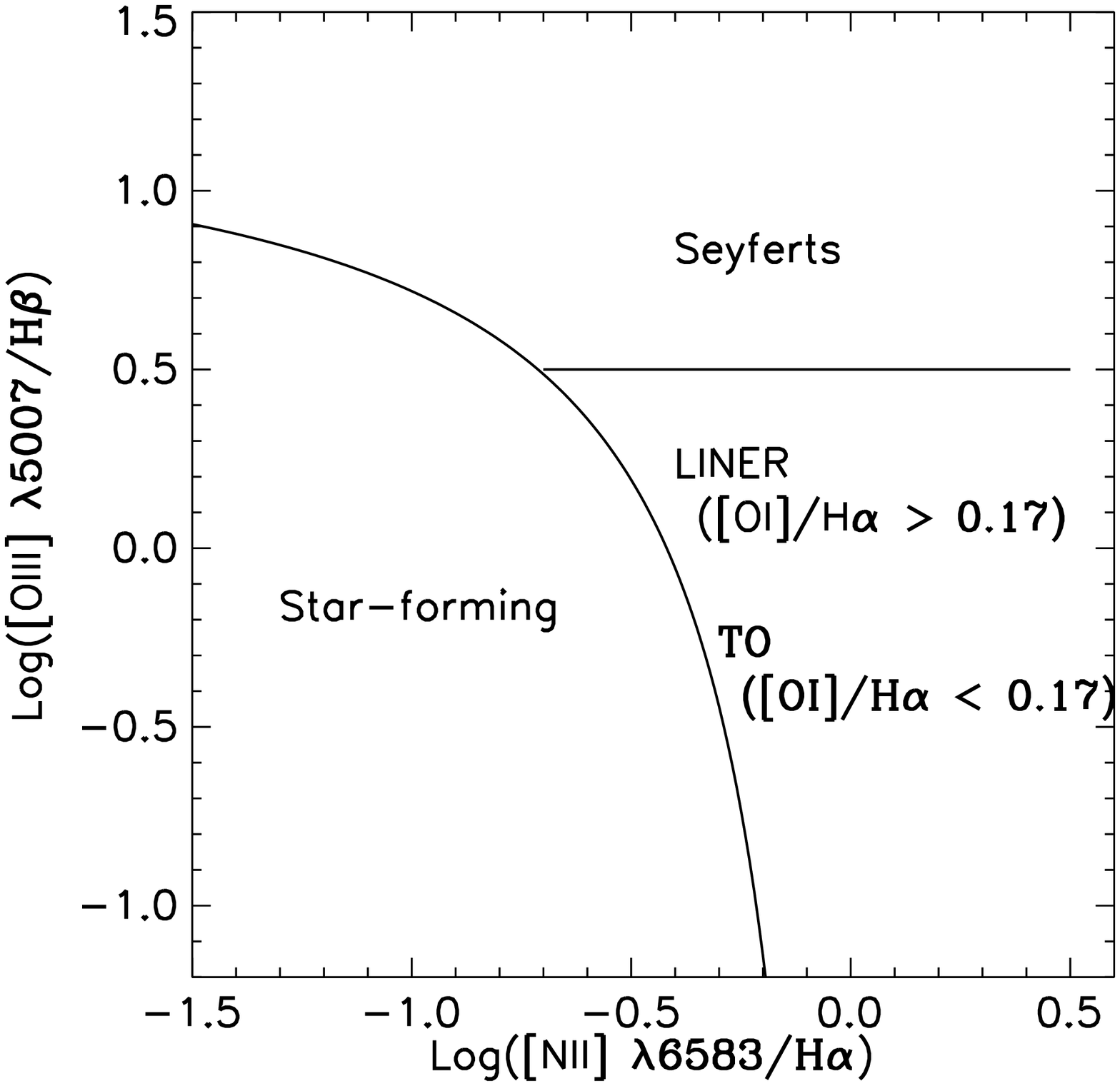}
\caption{This diagram shows the definition scheme described in \S\ref{sec:agn_mechanisms} and used in Fig.~\ref{fig:o2hb_3f}. We first employ the demarcation proposed by \cite{KauffmannHT03} to separate star-forming galaxies from other categories. Then we define Seyferts conventionally as galaxies with \oiiibw/\hb\ $>3$ \citep{ShuderO81,VeilleuxO87}. Last, we adopt the definition in \cite{HoFS97III} to separate LINERs and TOs using their \oiw/\hal\ ratio.}
\label{fig:definition}
\end{center}
\end{figure}

First we redivide our emission-line galaxies into four categories: star-forming galaxies, Seyferts, LINERs, and Transition Objects, using definitions similar to the conventions used in the literature. Our classification scheme is illustrated in Fig.~\ref{fig:definition}. We first employ the demarcation proposed by \cite{KauffmannHT03} to separate star-forming galaxies from other categories. Then we define Seyferts conventionally as galaxies with \oiiibw/\hb\ $> 3$. The remaining region belongs to LINERs and TOs. We adopt the definition in \cite{HoFS97III} to define LINERs as objects with $\oiw/\hal>0.17$ and TOs as objects with $\oiw/\hal<0.17$.

\begin{figure*}
\begin{center}
\includegraphics[angle=90,totalheight=0.35\textheight,viewport=0 0 360 760,clip]{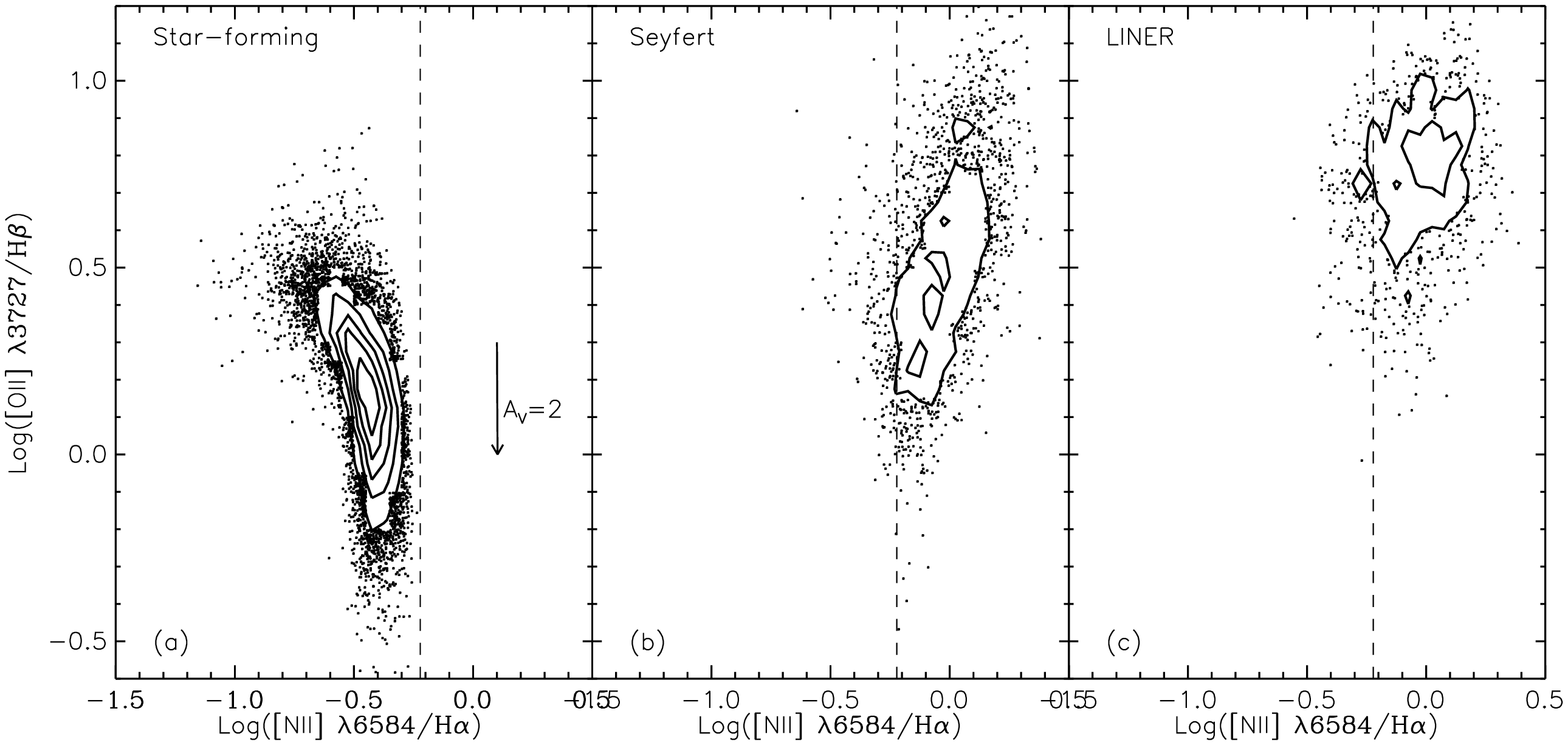}
\caption{
BPT-like diagrams showing \oiiw/\hb\ plotted against \niibw/\hal. The positions of star-forming galaxies (panel a), Seyferts (b) and LINERs (c) are shown separately. 
The effect of $A_V=2$ of extinction (for $R_V=3.1$) is indicated by the arrow in panel a.
There is very little vertical overlap between LINERs and star-forming galaxies in \oii/\hb\ ratio, which shows that star formation do not contribute substantialy to the \oii\ and \hb\ emission in LINERs. Note LINERs all fall in the high-\oii/\hb\ (equivalent to high-\oii/\hal) category. 
}
\label{fig:o2hb_3f}
\end{center}
\end{figure*}

Fig.~\ref{fig:o2hb_3f} shows a BPT-like diagram with the \oiiw/\hb\ ratio plotted against the \niibw/\hal\ ratio. A comparison between panel (a) and panel (c) shows clearly that LINER-like galaxies have no overlap with star-forming galaxies. If, for instance, star formation contributes substantially to the \oii\ emission in LINERs, it will also contribute to \hb\ in proportion according to a star-forming ratio. As a result, the \oii/\hb\ ratio and \nii/\hal\ ratio would be displaced towards the region occupied by star-forming galaxies. In fact, the LINERs have very little vertical overlap with star-forming galaxies in \oii/\hb\ ratio. Interestingly, Seyferts and Transition Objects do have huge overlap with star-forming galaxies, which is consistent with the picture that some of them might have substantial fractions of their \oii\ and \hb\ emission contributed by HII regions. Note LINER-like galaxies all fall in the high-\oii/\hb\ category, which is roughly equivalent to the high-\oii/\hal\ category. In fact, LINER-like galaxies have the highest \oii/\hb\ ratios observed. Thus it is impossible to produce LINER-like line ratios by combining Seyferts and HII region spectra.

It is logically possible that pure LINERs have more extreme ratios and star formation contribution brought them down to the current positions in the line ratio diagram. If that were true, the star formation rate would have to be remarkably uniform among all LINERs. Otherwise, we would see a spread of emission line ratios, producing an elongated distribution like those for Seyferts and Transition Objects. Thus this scenario is highly unlikely.

It is therefor clear that the high-\oii/\hal\ mode in the bimodality is mainly composed of LINERs and the line emission is produced by mechanisms other than star formation. 
However, the exact ionization mechanism for emissions with LINER-type line ratios is uncertain. 
The definition of LINER, as indicated by its name, refers only to emission 
regions around the nucleus of a galaxy, but LINER-like line ratios 
have also been observed in extended emission-line regions \citep[Graves et al. in prep]{Phillips86, 
HeckmanBvB89, Goudfrooij94, Zeilinger96, Goudfrooij99, Sarzi05}. Since SDSS 
spectra were taken with 3" fibers, which cover a region $\sim4{\rm kpc/h}$ 
in diameter at the median redshift of our sample, both the nuclei and possible 
extended emission regions in the galaxies would be included. 
Thus, we cannot tell whether the line emission in these 
red galaxies is associated with compact nuclear sources, extended 
regions, or both.

However, the occurence of LINER-like line ratios in
SDSS data is roughly consistent with results from the Palomar survey
\citep{FilippenkoS85,HoFS97III,HoFS97V},
which was conducted with a similar aperture (2"x4") but targeted very
nearby galaxies, therefore probing much smaller regions $\sim100$
parsec in diameter. Using that dataset, \citet{HoFS97V} found that
$\sim30\%$ of early-type galaxies host a LINER.
In the present SDSS sample, a similar fraction, namely
$>29\%$, of red galaxies have LINER-like emission line ratios.
We thus expect that if emission from extended
regions dominates the lines we observe, its line ratios and strengths must still correlate well with emission from the galaxies' nuclei.


Besides photoionization by a central non-stellar source, such as an 
accretion-powered system \citep{FerlandN83,HalpernS83,GrovesDS04II}, 
many ionization mechanisms have been proposed to explain the LINER- and 
TO-like optical emission in early-type galaxies:
(1) photoionization by the X-rays radiated as hot gas cools
\citep{VoitD90, DonahueV91, Kim89}, (2) heat transfer from hot gas to 
cooler gas \citep{SparksMG89}, (3) collisional 
excitation by shock waves \citep{HeckmanBvB89, DopitaS95}, 
(4) photoionization from some component of the stellar population, such as 
post-AGB stars \citep{dSA90,Binette94}.
We have compared the line ratios in the LINER-like SDSS galaxies with theoretical predictions of both shockwave models \citep{DopitaS95,DopitaS96} and the lowest-ionization Seyfert photoionization models \citep{GrovesDS04I, GrovesDS04II}. Both models produce line ratios roughly consistent with the observed ones. Detailed comparisons also require a careful extinction correction (given the possibility of patchiness) and metallicity measurements, which are beyond the limits of this paper. 
To reliably distinguish these different mechanisms, more information than just line 
ratio is needed, such as the spatial extent and kinematics of the 
emission-line regions, the spatial variation of line ratios, correlation with
X-ray, UV and radio properties, etc. 
It is also quite possible that many of these mechanisms are involved.
This is beyond the scope of the current paper, and as such  we do not try 
to answer this question here.

\section{Implications for Studies of Post-starburst Galaxies} 
\label{sec:implications}

\hal\ luminosity, the best optical indicator of star formation rate, is very 
difficult to measure at $z> 0.5$ due to atmospheric emission in the near infrared. 
Instead \oiiw\ luminosity is commonly used
\citep{DresslerG83, Zabludoff96,Hammer97, Hogg98, Dressler99, Poggianti99, Balogh99,RosaGonzalezTT02,TranFI03,Hippelein03, YangZZ04, TranFI04}.
However, we have shown above that the origin of \oii\ in most
red galaxies is not star formation. This could have strong consequences for
the definition of post-starburst galaxy samples, as they are approaching 
the red sequence and could suffer the same issues.

\begin{figure*}
\begin{center}
\includegraphics[totalheight=0.5\textheight]{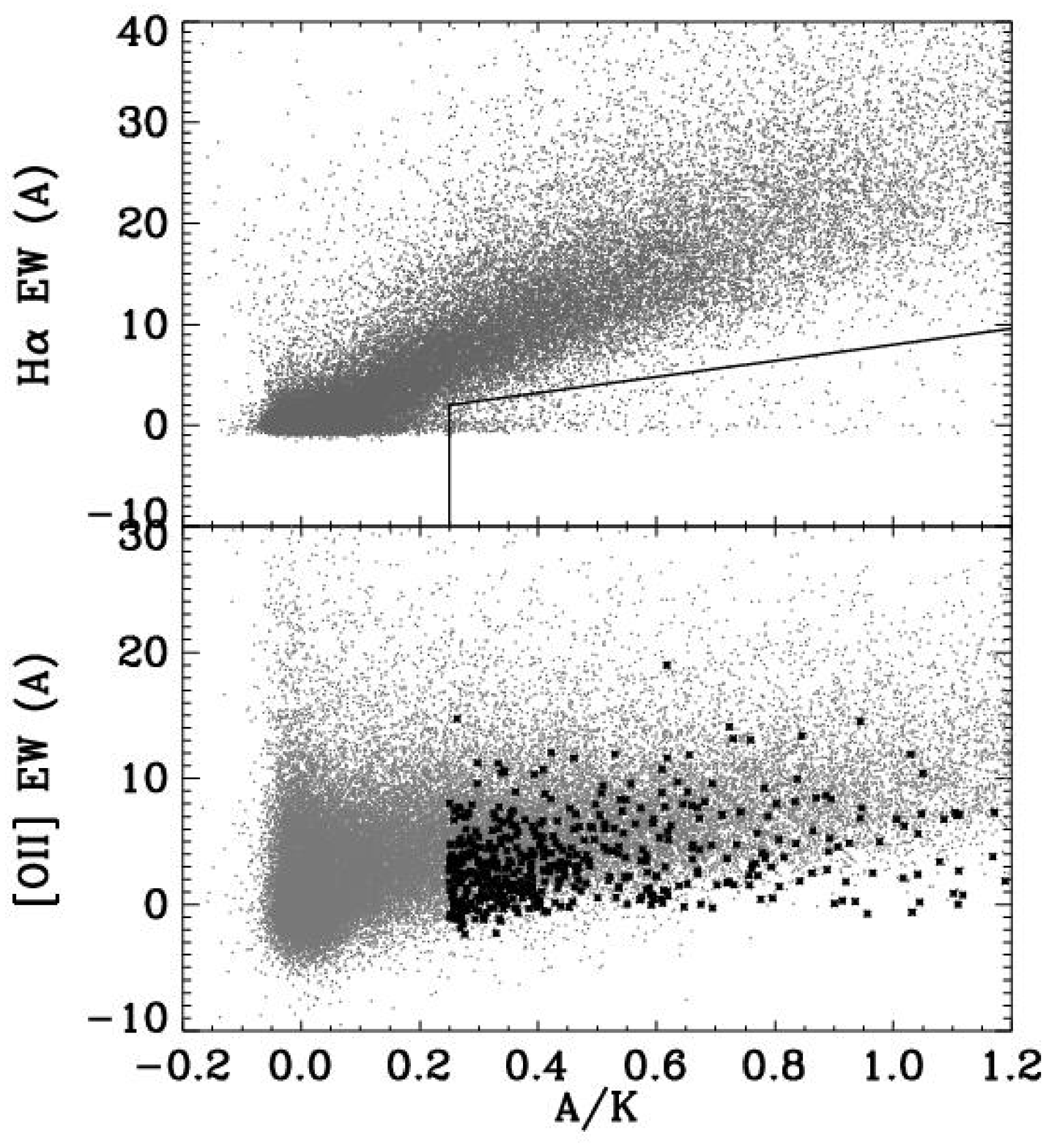}
\caption{Top panel: The distribution of \hal\ EW vs. A/K ratio for all 
galaxies with $0.05<z<0.2$ in the SDSS DR4 main galaxy sample that have both
\oii\ and \hal\ covered in their spectra (compare to Fig.3 in Quintero et al.).  
Post-starburst galaxies comprise a distinct population and stand out as the
horizontal spur with nearly zero \hal\ EW but relatively large A/K ratio. 
The empirical selection criteria are shown with the two solid lines.
Bottom panel: 
The distribution of \oii\ EW vs. A/K ratio for the same sample. The 
horizontal spur apparent in the top panel disappears. The dark points 
indicate the positions in the bottom panel of post-starburst galaxies 
selected in the top panel. These galaxies span a wide range in \oii\ EW and 
overlap with the region occupied by star-forming galaxies. 
}
\label{fig:ha_ak}
\end{center}
\end{figure*}

\cite{Quintero04} have demonstrated that post-starburst galaxies in the 
SDSS sample comprise a distinct 
population in a plot of \hal\ EW against A/K ratio\footnote{A/K ratio is the 
ratio between the coefficient of the young component to the coefficient of
the old component used in the linear decomposition of a galaxy spectrum, 
as described in Appendix \ref{sec:contsub}; a young stellar population will 
have high A/K ratio}. 
The \hal\ EW provides a measure of the current specific SFR in a galaxy, while 
the A/K ratio is linked to the average specific SFR over the past 1 Gyr;
comparing these two quantities indicates how 
the SFR is changing. In the top panel of Fig.~\ref{fig:ha_ak}\footnote{In this section, we plot all EW measurements regardless of the signal-to-noise of the detection, since galaxies without detectable emission are prime candidates to be post-starburst galaxies.}, we reproduce 
Fig.3 of Quintero et al. with our own measurements for all galaxies that are
within $0.07 < z < 0.1$ and have both \oii\ and \hal\ visible in their spectra.
For most galaxies, the two star formation indicators are highly correlated, as 
one would
expect. Post-starburst galaxies stand out as the horizontal spur
with nearly zero \hal\ EW but relatively large A/K, i.e., they have little or no
current SFR but large recent SFR. Alert readers will notice the difference 
between our A/K measurement from that of Quintero et al.'s; our linear 
decomposition is done after the subtraction of a continuum, while theirs 
is done without continuum subtraction; the templates used also differ.
In spite of this, the post-starburst population remains identifiable. 
We define post-starbursts as galaxies satisfying three criteria: 
${\rm A/K} > 0.25$, $\hal\ {\rm EW} < 8({\rm A/K})$ and $\hal\ {\rm EW} < 10{\rm \AA}$.

If we replace \hal\ by \oii, as shown in the bottom panel of 
Fig.~\ref{fig:ha_ak}, the post-starburst spur disappears.
Although they have minimal \hal\ emission, post-starburst galaxies
span a wide range of \oii\ EW. In a plot of \oii\ EW vs. A/K ratio, 
post-starburst galaxies identified using \hal\ overlap with the region occupied by star-forming galaxies, as seen 
in the lower panel of Fig.~\ref{fig:ha_ak}.

\begin{figure*}
\begin{center}
\includegraphics[angle=90,totalheight=0.38\textheight,viewport=0 15 370 700,clip]{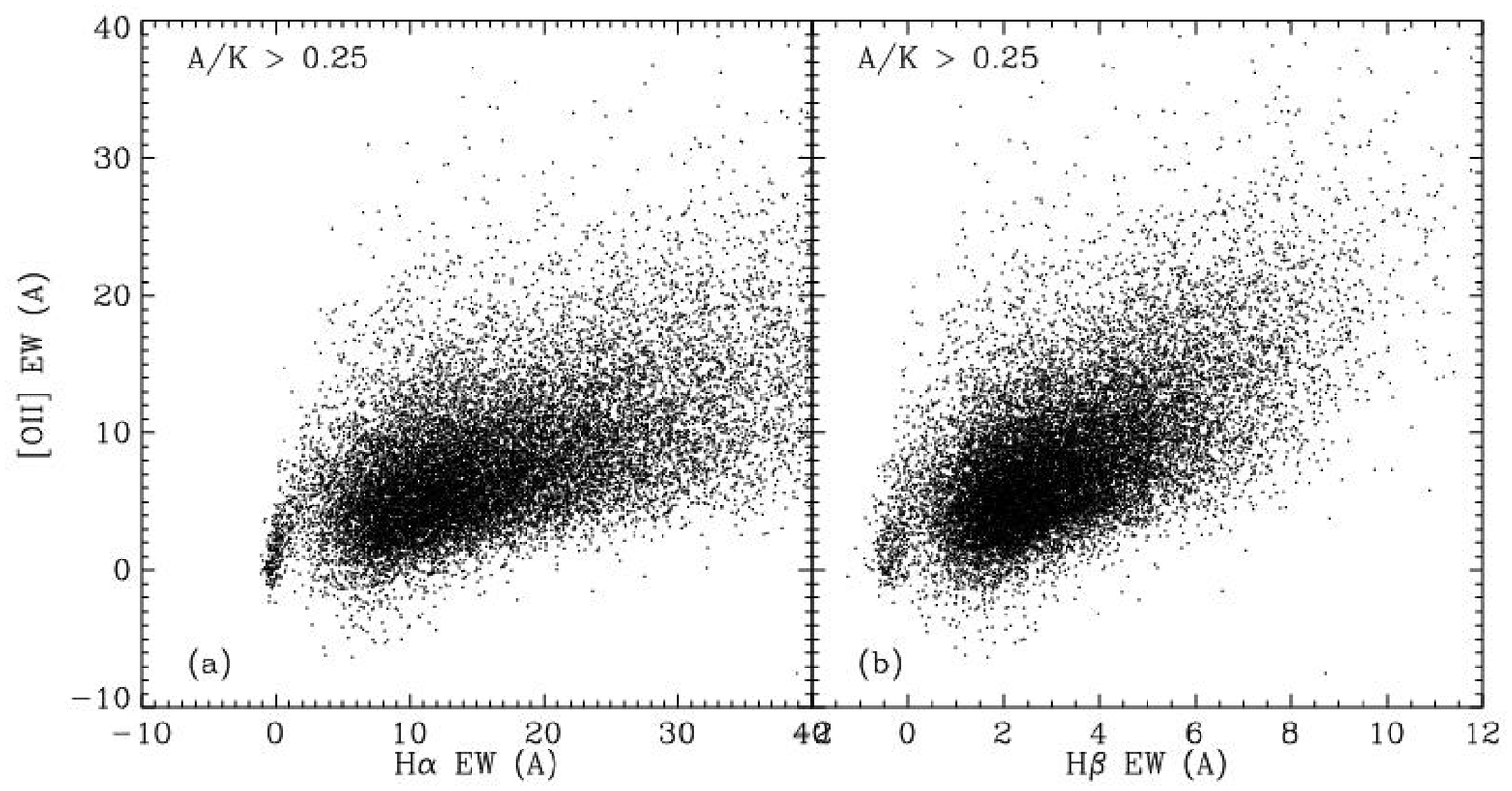}
\caption{Left: \oii\ EW vs. \hal\ EW distribution for galaxies with 
$A/K > 0.25$; i.e., those with relatively young stellar populations. Compare
to Fig.~\ref{fig:o2haew}a. However, here we include all EW measurements regardless of the signal-to-noise of the detection. Right: same galaxies as in left panel, but with \hb\ 
EW on the x-axis. In the \hb\ distriubtion, although EWs are smaller and measurement errors are larger
than for \hal, the separation between post-starburst and star-forming galaxies
remain apparent.}
\label{fig:akgt02}
\end{center}
\end{figure*}

The left panel of Fig.~\ref{fig:akgt02} shows more directly what causes 
the difference between the two panels of Fig.~\ref{fig:ha_ak}. 
This figure replots Fig.~\ref{fig:o2haew} but using only galaxies 
with an A/K greater than 0.25. This corresponds to the A/K cut 
\cite{Quintero04} used to define their post-starburst (K+A) sample; 
in this plot, we include all galaxies regardless of \hal\ EW. 
As expected, most galaxies on the star-forming sequence from 
Fig.~\ref{fig:o2haew} remain, while most red galaxies are excluded. However, 
there is a small group of points forming a vertical spike near zero 
\hal\ EW; these are post-starburst galaxies having high A/K and 
little or no \hal\ emission.  There is a gap in \hal\ that
naturally separates this post-starburst population from star-forming 
galaxies. However, there exists no corresponding gap in \oii\ EW. 

In fact, roughly 50\% of post-starburst (K+A) galaxies identified using \hal\ belong to either the
high-\oii/\hal\ population or the \oii-only population defined in \S\ref{sec:comp}; the remainder are classified as quiescent. Although they show
minimal \hal\ emission, they have LINER/AGN-like line ratios, and can have
appreciable \oii\ emission. Thus, \hal\ can be used as a universal 
star formation indicator for most cases with little contamination. However,
in red and post-starburst galaxies, LINER/AGN can possess \oii\
lines as strong as those in star-forming galaxies. 

For this reason, using \oii\ to identify post-starbursts is problematic.
Adopting a low \oii\ EW cut would make the 
selection heavily incomplete, while a high cut would bring huge
contamination from star-forming galaxies. In addition, differing levels of 
measurement 
errors will cause varying degrees of contamination. This could well explain the 
drastically different post-starburst abundances found in some previous 
studies \citep{Dressler99, Poggianti99, Balogh99}.
In the volume-limited sample, 66\% of post-starburst galaxies have detectable \oii\ emission; 57\% have \oii\ EW greater than 2.5\AA. Previous studies which used an \oii\ EW cut of this level could be missing half of all the post-starbursts.

Unlike \oii, \hal\ appears to provide a generally reliable indication of the
presence of star formation in the absence of a full BPT diagnosis. 
However, at $z > 0.4$, \hal\ is difficult to measure due to strong 
atmospheric emission and declining instrumental sensitivity. \oii\ has been 
used as a proxy for measuring SFR at redshifts up to $z\sim 1.4$, but as 
shown above, it is problematic for red and post-starburst galaxies. In 
its place, we suggest the use of \hb, which is available out to higher
redshifts than \hal, but is much more reliable for red galaxies than \oii. 
In the DEEP2 Galaxy Redshift Survey \citep{Davis03}, \hb\ can be measured up to 
$z\sim 0.9$, enabling a direct comparison with low-z studies. Although 
it is much weaker than \hal, more affected by dust, and has a deeper stellar 
absorption component to remove, \hb\ behaves similarly to \hal\ in SDSS 
data, as shown in the right panel of Fig.~\ref{fig:akgt02}. 
In a future paper, we will use \hb\ 
to identify post-starburst galaxies in DEEP2 and compare with low-z results. 
Finally, for completeness, we also give an empirical bimodality demarcation 
for \oii-\hb:
\begin{equation}
{\rm EW}(\oii) = 18{\rm EW}(\hb) -6 .
\end{equation}

\subsection{The Nature of Post-starburst emission}

Tradiationally, post-starburst galaxies have been defined as galaxies with
no detectable line emission. \cite{Quintero04} enlarged
the definition to include galaxies with nonzero \hal\ emission but too weak
compared to star-forming galaxies with a similar fraction of young stars 
(similar A/K ratio, see Fig.~\ref{fig:ha_ak}).
As shown above, these galaxies can have rather strong \oii\ lines, even when
\hal\ is not detectable (consistent with zero). In the volume-limited sample, 78\% of post-starburst galaxies have at least one emission line detected at $>2\sigma$ significance. What is the nature of this emission? Does it indicate residual star formation? 

\begin{figure}
\begin{center}
\includegraphics[totalheight=0.48\textheight]{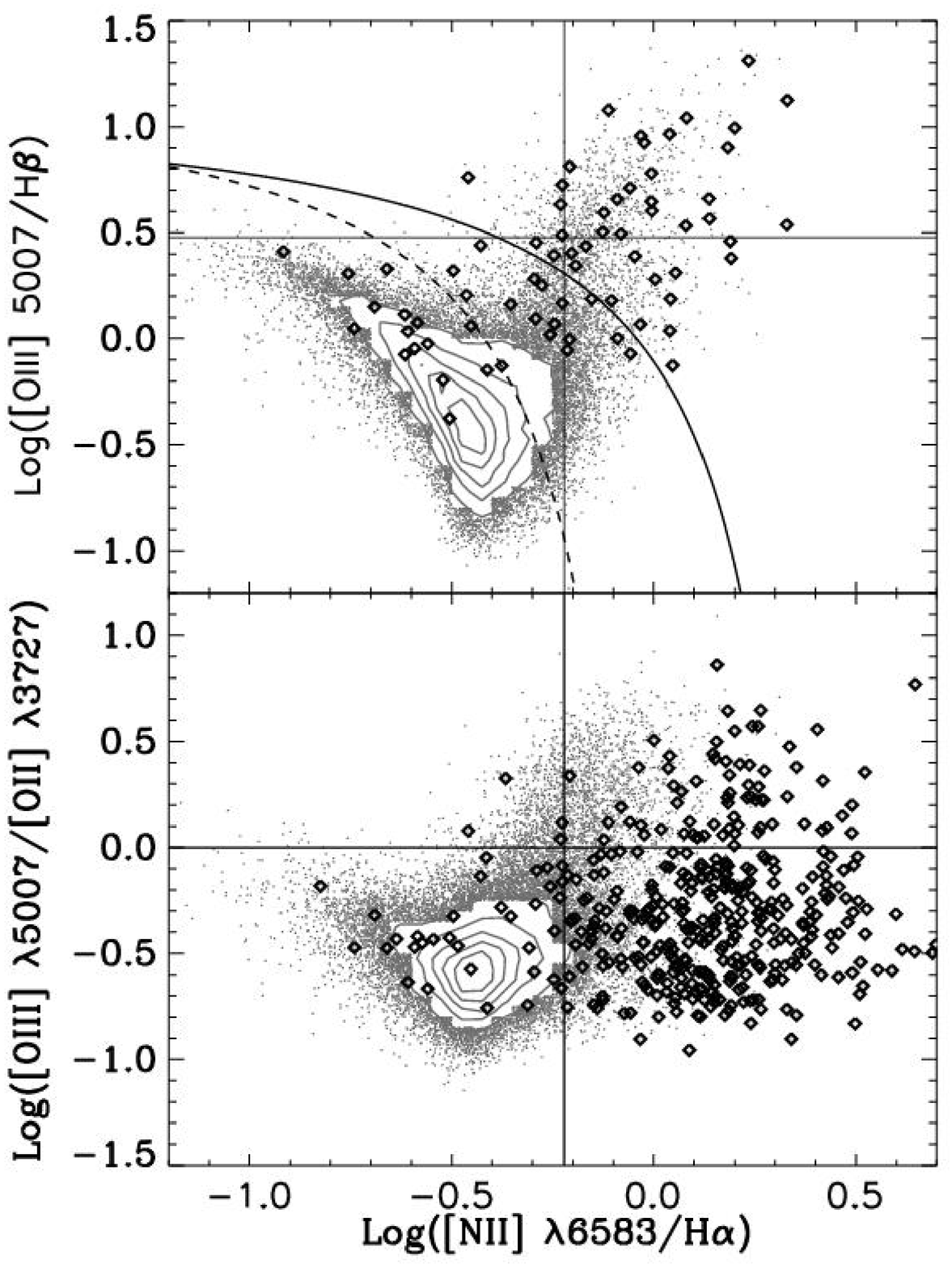}
\caption{Line ratio diagnostic diagrams for galaxies with $A/K > 0.25$ in the magnitude limited sample: \oiiibw/\hb\ vs. \niibw/\hal\ (top) and \oiiibw/\oiiw\ vs. \niibw/\hal\ (bottom).
All galaxies plotted here have at least 2$\sigma$
detections on the emission lines involved. Post-starburst galaxies are 
highlighted with the diamond symbol; they mostly have LINER-type line ratios. The number of post-starburst galaxies plotted in the top panel is much smaller because \hb\ is more difficult to detect than the strong lines used in the bottom panel.}
\label{fig:ak_bpt}
\end{center}
\end{figure}

For post-starbursts that have reliable line detections, we can again
employ BPT diagrams to study the nature of their emission. 
Fig.~\ref{fig:ak_bpt} shows two BPT diagrams for galaxies with A/K greater 
than 0.25 in the magnitude-limited sample ($0.05 < z < 0.1$), i.e.,  
only galaxies which contain young stellar populations. 
As before, we require a minimum detection of 2$\sigma$ 
on all lines, which excludes some fraction of the post-starburst galaxies from the diagrams depending on the lines used. For example, about 1/3 of all post-starbursts are included in the bottom panel but only 5\% in the top panel. 
As shown in this figure, most post-starburst 
galaxies in which all emission lines involved are 
positively detected show AGN-type line ratios and cover all 
categories of AGN -- LINER, TO, and Seyfert. Only 6\% of these 
post-starbursts show line ratios characteristic of star-forming HII regions,
which also turn out to be preferentially high-\hal\ EW cases. 

Those post-starbursts that cannot be plotted on the BPT diagram mostly belong
to the high-\oii/\hal\ group, the \oii-only group, or the quiescent group. Of the $\sim20\%$ low-\oii/\hal\ and \hal-only galaxies, the majority have \nii/\hal$>0.6$ suggesting a non-starforming origin for the emission lines.

We can break down the classifications for all post-starbursts following the scheme used in Table~\ref{tab:red_breakdown} and use the \oii-\hal\ demarcation and \nii/\hal\ ratio to discriminate between star-forming galaxies and other groups. We find only 5\% of post-starbursts in the volume limited sample have emission lines dominated by residual star formation.
Instead, the emission in most post-starbursts has AGN-like line ratios.
Recently, much evidence has suggested a connection between AGN 
and post-starbursts \citep{Brotherton99, CanalizoS01, KauffmannHT03, 
Heckman04, Ho05}. Do AGN play a role in quenching star formation? 
The fact that post-starbursts have AGN-like line ratios is consistent 
with this picture.

Conservatively speaking, without knowledge of the spatial distribution of 
emission and information from other wavebands, we cannot conclude definitely 
that these are true AGNs. 
Unlike most red, elliptical galaxies, post-starburst galaxies have 
young stellar populations, which allow at least one more candidate 
excitation mechanism for LINER-type line ratios aside from those described 
above in \S\ref{sec:agn_mechanisms}.
\cite{Taniguchi00} point out that hot planetary nebula nuclei formed 
100-500Myr following a starburst could provide the ionizing spectra necessary 
to form LINER-type emission lines.  
Further observations of post-starburst galaxies in other wavebands and more 
detailed modeling should provide a final answer to this question. If AGNs 
are in fact ubiquitous among post-starburst galaxies, it may provide a hint 
that AGN played a role in the quenching of their star formation.

\section{Summary}\label{sec:summary}

We have measured the fluxes and EWs of \oii, \hal\, and several other emission 
lines for $\sim 
300,000$ galaxies in the SDSS DR4 main galaxy sample after careful subtraction
of the stellar continua. From the comparison 
between \oii\ and \hal, combined with a variety of line ratio diagnostics, we 
have 
investigated the origin of line emission in red galaxies and the implications
for post-starburst galaxy studies. Our main conclusions are as follows:

\begin{enumerate}
\item

Galaxies display a bimodality in \oii/\hal\ ratio, which corresponds closely 
to the bimodality in their rest-frame colors. In blue galaxies, the 
\oii/\hal\ ratio generally matches the expectation for star-forming HII 
regions. However, in most red galaxies, the \oii/\hal\ ratio is unusually high 
and difficult to reconcile with predictions for star-forming HII regions. 

\item 
About 52\% of all red galaxies have detectable line emission; 38\% have detectable \oiiw\ emission. 
More than 29\% of all red galaxies have line ratios characteristic of LINERs, while less than 17\% are TOs or Seyferts, and $\sim6\%$ have lines dominated by star formation (probably dusty starbursts and edge-on spirals).
Further information about the spatial distribution of 
emission in these galaxies and/or X-ray, UV and radio observation is necessary 
to firmly identify the origin of the line emission in each case. 

\item 
LINER-like galaxies make up the high-\oii/\hal\ mode in the \oii/\hal\ bimodality. Other than modest differences in the luminosity range spanned, they are essentially indistinguishable from quiescent galaxies in the color-magnitude-concentration space. The combination of LINER-like and quiescent galaxies defines a uniform red sequence in color-magnitude-concentration space, which can be effectively selected by a simple division in the \oii-\hal\ EW diagram.
The remaining red galaxies, those with low \oii/\hal\ ratio, are mostly TOs, dusty-starforming galaxies, and a small fraction of Seyferts. This group also have lower galactic concentration and slightly bluer color than the LINER/Quiescent red sequence.

\item
Post-starburst galaxies, identified in the SDSS dataset using a lack of \hal\ 
emission to indicate that star formation has ceased, often
exhibit significant \oii\ EW. Their positions in the \oii-\hal\ plot and 
BPT diagrams are the same as for Seyferts, LINERs and TOs, which suggest they 
may harbor AGNs. Less than 5\% of this post-starburst sample show 
evidence for residual star formation.

\item
\oii\ emission in red galaxies and in post-starburst galaxies (i.e., in cases where it is not induced by star-formation) can be as strong as in star-forming
galaxies. Locally at least, \oii\ can only be used as a star-formation 
indicator for blue galaxies.

\item
More than half of all the post-starbursts have detectable \oii\ emission.
Post-starburst samples defined using \oii\ as a star-formation indicator 
will be very incomplete, especially if AGN/LINER rates and intensity were 
higher in the past. We recommend using \hb\ as an alternative for 
high-redshift studies.

\end{enumerate}

\acknowledgements
We would like to thank David Hogg and Alex Quintero for discussions that 
inspired this work. We also would like to thank Michael Blanton and David 
Schlegel for help in using the SDSS data.
RY thanks Guinevere Kauffmann, Luis Ho, Robert Kennicutt, Timothy Heckman,
Daniel Eisenstein, Alex Filippenko, and Ricardo Schiavon for helpful discussions. RY also 
thanks Michael Cooper, Darren Croton, Brian Gerke and Genevieve Graves for careful reading of this manuscript.
Finally, we would like to thank the referee for helpful comments which significantly improved this paper. 
 JN acknowledges support from NASA through the Hubble Fellowship grant 
 HST-HF-01165.01-A awarded by the Space Telescope Science Institute, 
 which is operated by the Association of Universities for Research in 
 Astronomy, Inc., for NASA, under contract NAS 5-26555.
 SMF would like to acknowledge the support of a Visiting Miller 
 Professorship at UC Berkeley.
The project was supported in part by the NSF grants AST00-71198 and 
AST00-71048.
This research made use of the NASA Astrophysics Data System, and employed
open-source software written and maintained by David Schlegel, 
Douglas Finkbeiner, and others.

 Funding for the Sloan Digital Sky Survey (SDSS) has been provided 
 by the Alfred P. Sloan Foundation, the Participating Institutions, 
 the National Aeronautics and Space Administration, the National 
 Science Foundation, the U.S. Department of Energy, the Japanese 
 Monbukagakusho, and the Max Planck Society. The SDSS Web site is 
 http://www.sdss.org/.

 The SDSS is managed by the Astrophysical Research Consortium (ARC) 
 for the Participating Institutions. The Participating Institutions are 
 The University of Chicago, Fermilab, the Institute for Advanced Study, 
 the Japan Participation Group, The Johns Hopkins University, Los 
 Alamos National Laboratory, the Max-Planck-Institute for Astronomy 
 (MPIA), the Max-Planck-Institute for Astrophysics (MPA), New Mexico 
 State University, University of Pittsburgh, Princeton University, the 
 United States Naval Observatory, and the University of Washington. 

\appendix
\section{Stellar continuum subtraction} \label{sec:contsub}
To get an accurate measurement of the emission line fluxes and equivalent widths, 
we need to subtract 
the stellar component underneath. All Balmer lines have a broad absorption 
component, while the continuum underlying \oii\ has many narrower absorption features. 
The best way of removing the stellar component is often to fit the entire 
spectrum with a stellar population synthesis model. 
Here, we trade off this freedom for robustness and efficiency. 

Instead of doing full population synthesis modeling, we 
decompose the spectra into two components: an old stellar population 
and a young stellar population.
Similar methods have been used in modeling the spectra of E+A galaxies, 
such as in \cite{Quintero04}, who employed an A-star spectrum and a composite 
luminous red galaxy spectrum as templates. As pointed out by \cite{DresslerG83}, 
a linear combination of the two components can simultaneously reproduce 
both the broad spectral continuum and the depths of narrow features for
post-starburst galaxies, over the range of 3400 to 5400 angstroms. 
For galaxies other than post-starbursts, the broadband continuum is not 
always matchable by this method, but the depths of the narrow 
features can always be well fitted by a combination of these two components;
this is sufficient for our purposes. Thus, we model the continuum-subtracted 
spectra by the combination of the two templates, each continuum-subtracted 
in the same way. The broadband continuum subtracted is determined by the 
inverse variance-weighted mean of the spectrum in a 300\AA\ window. The 
method works well for the regions around all emission lines used here: viz., \oiiw, \hb, \oiiibw, \oi, \hal, and \niibw; see \S\ref{sec:testsub}. 

The coefficients of the two templates are obtained from a linear fit over 
the restframe wavelength range $3600{\rm \AA} < \lambda < 5400{\rm \AA}$.
Before fitting, both templates are convolved with a Gaussian kernel with 
a dispersion equivalent to the combination of the instrumental dispersion and 
the velocity dispersion of the galaxy being fitted in quadrature.
The fitting is weighted by the inverse variance in the observed 
spectrum. We mask regions around emission lines using adaptive 
windows set by the velocity dispersion measurement of each galaxy.

Spectra that cover a total of less than 900\AA\ (3 times the broadband
continuum subtraction window) in the observed frame or less than 200\AA\
out of the restframe
wavelength range from 3600\AA\ to 5400\AA\ are excluded from the samples used in this paper, in order to ensure good removal of stellar continuum features.

\subsection{The Templates}
Our young stellar population template is given by a model spectrum built 
using the new population synthesis code of \cite{BC03}. The template spectrum 
is taken at 0.3 Gyr after a starburst
with a constant star formation rate lasting 0.1 Gyr with solar 
metallicity. This template is used instead of an A-star spectrum 
because it is a more physical choice. Moreover, it includes
all the spectral features produced in intermediate and low mass stars
which also exist in a starburst. This particular choice of spectrum also 
provides a
relatively good subtraction of the continuum features for galaxies with ongoing
star formation.

The old stellar population template is also built using \cite{BC03} code.
The model used is a 7Gyr old simple stellar population with solar metallicity. 
One could have instead used a purely empirical template, such as the SDSS 
Luminous Red Galaxy coadded spectrum \citep{EisensteinHF03}. This spectrum 
closely resembles our synthesized old stellar population spectrum if the 
broadband continuum is subtracted: almost every feature in the two spectra 
matches. However, such a spectrum appears to include the sort of line emission 
we are looking for; e.g., it has an \oii\ EW of $\sim 2$\AA. Thus, we avoided
using it as a template.



\subsection{Testing the subtraction} \label{sec:testsub}
We have tested our subtraction of the stellar continuum using a set of population 
synthesis models and real spectra. Except for models with very recent 
bursts (younger than 200 Myr), the residuals of the fits are sufficiently small
(the residual EWs in \hb\ are within $\pm0.4$\AA\ and those in \hal\ are within 
$\pm0.7$\AA). For recent bursts the discrepancy occurs because such galaxies
are likely to be dominated by B stars, which have narrower Balmer absorption 
lines than A dwarfs. A characteristic residual is therefore apparent for these 
galaxies. However, we have found this sort of feature only very rarely ($\lesssim1\%$)
in visual inspection of residuals from a subset of the spectra used here.
Since this paper focuses on emission lines in
red galaxies, the poor residuals for extremely young bursts do not affect 
our conclusions. In some cases, there could also be systematic residuals around
the Ca K line, which is a metallicity-sensitive feature. Since this feature 
does not overlap any of the emission lines we measure, it does not affect 
our results. 

\begin{figure}
\begin{center}
\resizebox{7in}{!}{\includegraphics{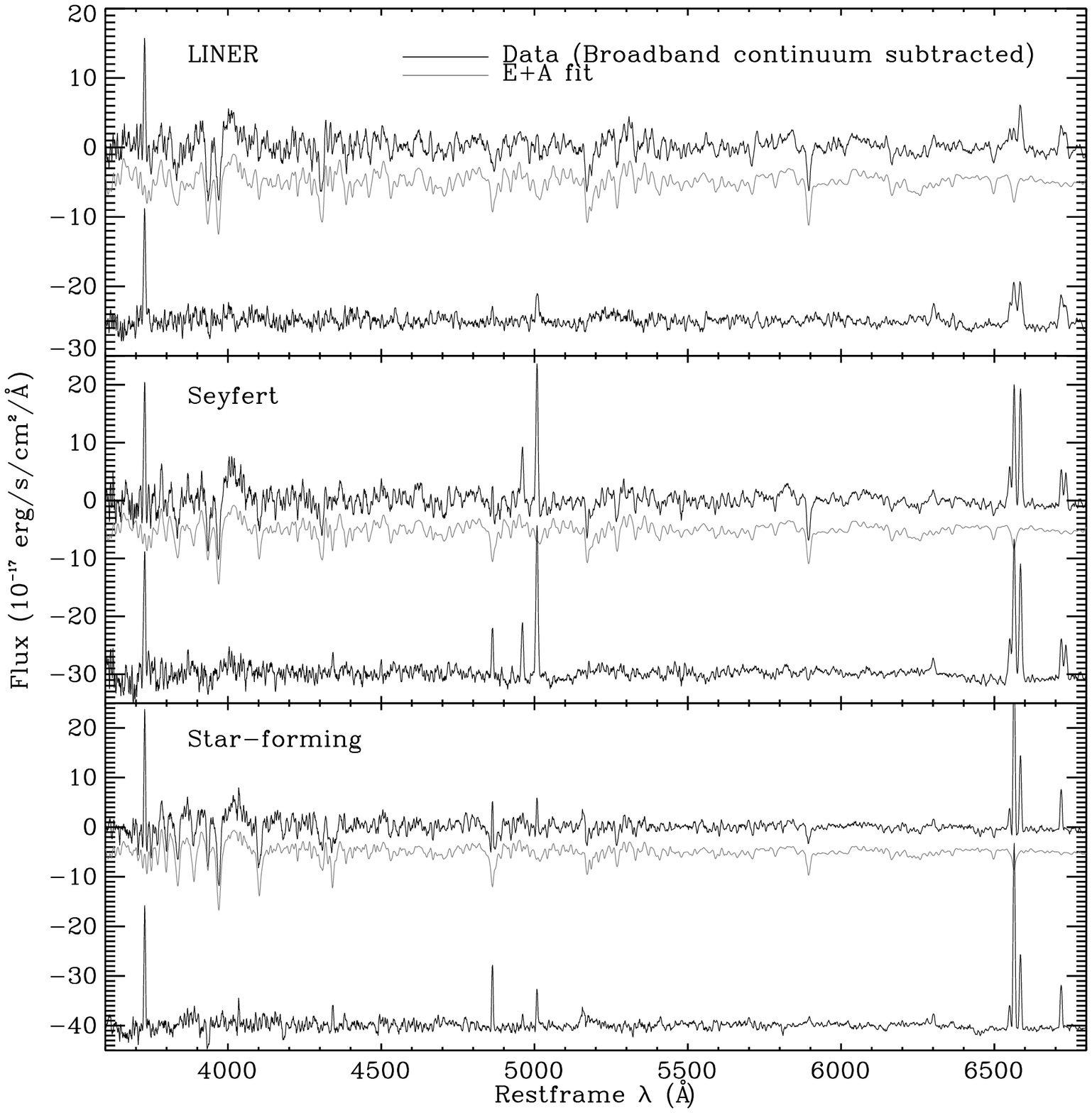}}
\caption{Sample spectra illustrating the accuracy of our stellar continuum subtraction. 
These spectra are selected to have a S/N ratio near the median value
in the volume-limited sample. The fit and residual are offset for clarity. 
Weak emission lines are more easily seen in the residual than in the original 
spectra, such as \oiiibw\ in the top panel and \oiw\ in all three spectra. 
Also worth noting is that the different line ratios show the different origins of the 
line emission, as discussed in \S\ref{sec:comp} and \S\ref{sec:agn}. }
\label{fig:subtraction}
\end{center}
\end{figure}

Fig.~\ref{fig:subtraction} shows three examples of the stellar continuum 
subtraction. They are selected to have a S/N ratio near 
the median value in the sample. The residual of the subtraction is shown at the
bottom of each panel. Most stellar features are successfully subtracted with small 
residuals. In high S/N spectra where residuals are not dominated by noise, the median root-mean-square of the residual around each line is less than 2\% of the continuum level except around \oii, where residuals approach 10\% (this is an uppper limit, as noise in the spectra is higher at shorter wavelengths).

\section{Line measurements} \label{sec:line}
We measure the emission line flux by two methods: flux summing and Gaussian 
fitting. Table \ref{tab:line} lists our line definitions. For each emission 
line, we define a center passband in which we measure the line flux, and 
two sidebands in which we measure the continuum flux. We use these same wavelength ranges to establish which spectra are
well-measured in the vicinity of \hal\ and \oii, defining our sample.
We require that at least 83\% of all pixels in the center passband
and 17\% of the pixels in each sideband be well-measured for both \oii
and \hal.  In order to be considered good, we require both a pixel and
its four adjoining pixels to have inverse variance above the greater of
$10^{-10}$ and one-ninth the median inverse variance for the spectrum.

First, the stellar continuum is reconstructed via our linear fit with two 
templates. Although the fit is performed with the broadband continuum 
subtracted templates, in reconstruction, we use the original templates with
the two linear coefficients from the fit. This reconstructed stellar continuum
has correct depths for narrow features but offset overall amplitude.
We then add a constant to it to match the median of the observed 
flux in the two sidebands, respectively for each emission line considered.
Subtracting off this stellar continuum from the observed spectrum then leaves us 
with an emission-line only spectrum. The line flux is measured by integration 
in the center passband.  

In addition to measuring total flux from simple flux-summing in a window, 
we also apply a Gaussian fitting technique. Here \hb, \oiiibw, and \oiw\ 
are each modeled as single Gaussians, where the line amplitude, dispersion 
and central wavelength are fit simultaneously by a Levenberg-Marquardt 
least-squares method with realistic boundaries set on
the last two parameters. \oiiw\ is fitted
as a doublet with the wavelength ratio fixed and dispersion shared
between the two lines. Because of their proximity, \hal\ and the \niiw\ doublet 
are fit simultaneously with three Gaussians 
with wavelength ratios fixed and dispersion shared.
We have checked the line fluxes obtained from flux-summing against those from
Gaussian fitting. Good agreement is achieved for lines with detectable EWs.  For instance, the standard deviation of the fractional difference in \hal\ flux is about 7\% for galaxies with \hal\ EW around 1\AA; that of \oii\ flux is about 16\% at EW(\oii)$\sim 4{\rm \AA}$. Systematic differences occur near zero flux, as the Gaussian fitting routine used tends to avoid low amplitudes, fit a broad Gaussian to the noise and overestimate the flux. For this reason, we use the flux-summing measurements for all results presented in this paper.

As described in \S\ref{sec:subsample} and illustrated in Fig.~\ref{fig:oii_dist}, the measured EW distributions of galaxies that fall to the left of the demarcation in the \oii-\hal\ EW diagram can be modeled well by the combination of a Gaussian distribution centered near zero and a log-normal distribution starting from the peak of the normal distribution; these two components correspond very closely to the quiescent population and the LINER-like population, respectively. We make use of this fact to measure zero point offsets for our emission line measurements, as presumably the true EW for quiescent galaxies is zero or extremely close to it. Table~\ref{tab:zeropoint} lists the zero point determined by this method for each emission line used in this paper and the $\sigma$ of the Gaussian distribution of that line's EW for quiescent galaxies. The latter gives a measure of the line detectability. We have subtracted the zero point derived by this method from all the EW values used in this paper. The $\sigma$'s of the Gaussian distributions for \hal\ and \hb\ suggest the EW errors for these two lines are too low by 33\%, perhaps due to the difficulty of subtracting broad Balmer absorption features.  We thus multiply the errors used for these lines in the paper by 1.5 when determining significances, etc.  For all other lines, the Gaussian width is in good agreement with the estimated errors.

\begin{table}
\begin{center}
\begin{tabular}{ccccc}
  Line  & Zeropoint (\AA) & $\sigma$ (\AA) \\ \hline
  \oiiw & 1.45 & 1.56 \\  
  \hb &  0.318 & 0.231 \\
  \oiiibw &  0.54 & 0.35 \\
  \oiw &  0.119 & 0.238\\
  \hal & 0.487 & 0.349 \\
  \niibw & 0.195 & 0.312 \\
\end{tabular}
\caption{
For each emission line used in this paper, we list the zero point offset of the original EW measurements and the standard deviation ($\sigma$) in EW for galaxies whose EW is consistent with zero. These values have been determined from the Gaussian + log-normal decomposition of the EW distribution for galaxies to the left of the demarcation in \oii-\hal EW diagram.
}
\label{tab:zeropoint}
\end{center}
\end{table}

We have also checked our EW measurements against those measured by Christy 
Tremonti \citep[see][]{TremontiHK04}. In almost every case, any zeropoint 
offset 
is small enough to be ignored: the difference between the two measurements are 
of order of the error estimates. However, for \hb, our EW is about 1\AA\ 
smaller than \cite{TremontiHK04} when \hb\ is greater than 10\AA. This is of 
little concern, because our main interest in this paper centers around red 
galaxies with little \hb\ emission, where 
the zeropoint offset is less than 0.1\AA. The difference is likely to be due to 
differences in templates and methods for stellar continuum subtraction; galaxies 
with strong star formation will typically exhibit both strong \hb\ emission and
significant stellar \hb\ absorption features.

\begin{table}
\begin{center}
\begin{tabular}{ccccc}
  Index  & Bandpass & Blue Sideband & Red Sideband & Reference \\ \hline
  \oiiw & 3716.3 - 3738.3 & 3696.3 -3716.3 & 3738.3 - 3758.3 & \cite{FisherFF98}\\
  \hb & 4857.45-4867.05 $^{a}$ & 4798.875-4838.875 & 4885.625 - 4925.625 & \cite{FisherFF98}\\
  \oiiibw & 4998.2-5018.2 & 4978.2-4998.2 & 5018.2-5038.2 & \\
  \oiw &  6292.0-6312.0 & 6272.0-6292.0 & 6312.0-6332.0 & \\
  \hal & 6554.6 - 6574.6 & 6483.0 - 6513.0 & 6623.0 - 6653.0 &\\
  \niibw & 6575.3 - 6595.3 & 6483.0 - 6513.0 & 6623.0 - 6653.0 & 
\end{tabular}
\tablenotetext{a}{The central bandpass is modified to cover only the middle 1/3 
 of the window in the \citet{FisherFF98} definition. The integrated flux 
 in the modified window agrees better (i.e., has less scatter) with the 
 Gaussian fitting results than the flux in the wider, original window does.}
\caption{Window definitions for the emission lines we measure.}
\label{tab:line}
\end{center}
\end{table}


\end{document}